\RequirePackage{lineno}
\documentclass[floatfix,twocolumn,prd,showpacs,preprintnumbers,amsmath,nofootinbib,amssymb,superscriptaddress]{revtex4-1}
\usepackage{mathtools, cuted}
\usepackage{lipsum}
\usepackage[utf8]{inputenc}
\usepackage{graphicx}
\usepackage{dsfont}
\usepackage{mathrsfs}
\usepackage{bm}
\usepackage{color}
\usepackage[colorlinks,citecolor=blue,urlcolor=blue,linkcolor=black]{hyperref}
\usepackage{braket}
\usepackage{placeins}
\usepackage{multirow}
\usepackage{hyperref}
\usepackage{float}
\usepackage{booktabs}
\usepackage{soul}
\usepackage{float}
\usepackage{tikz,pgfplots,pgfplotstable}
\usepgfplotslibrary{groupplots,dateplot}
\usetikzlibrary{patterns,shapes.arrows}
\pgfplotsset{compat=newest}
\usepgfplotslibrary{fillbetween}
\usetikzlibrary{external}
\usepackage[nameinlink,capitalize]{cleveref}

\newcommand{\dd}{\mathrm{d}}
\newcommand{\Jpsi}{J/\psi}
\usepackage{blindtext}

\begin{document}

\title{Charm and beauty in the deconfined plasma from quenched lattice QCD}

\author{H.-T. Ding}
\affiliation{Key Laboratory of Quark and Lepton Physics (MOE) and Institute of
    Particle Physics, \\
    Central China Normal University, Wuhan 430079, China}
\author{O. Kaczmarek}
\affiliation{Key Laboratory of Quark and Lepton Physics (MOE) and Institute of
    Particle Physics, \\
    Central China Normal University, Wuhan 430079, China}
\affiliation{Fakult\"at f\"ur Physik, Universit\"at Bielefeld, D-33615 Bielefeld,
    Germany}
\author{A.-L.~Lorenz}
\affiliation{Fakult\"at f\"ur Physik, Universit\"at Bielefeld, D-33615 Bielefeld,
    Germany}
\author{H. Ohno}
\affiliation{Center for Computational Sciences, University of Tsukuba, Tsukuba, Ibaraki 305-8577, Japan}
\author{H. Sandmeyer}
\affiliation{Fakult\"at f\"ur Physik, Universit\"at Bielefeld, D-33615 Bielefeld,
    Germany}
\author{H.-T. Shu}
\thanks{Current address: Institut f\"ur Theoretische Physik, Universit\"at   Regensburg, D-93040 Regensburg,
    Germany.}
\affiliation{Fakult\"at f\"ur Physik, Universit\"at Bielefeld, D-33615 Bielefeld,
    Germany}


\begin{abstract}
We present continuum extrapolated results of charmonium and bottomonium correlators in the vector channel at several temperatures below and above $T_c$. The continuum extrapolation jointly performed with the interpolations to have physical values of $\Jpsi$ and $\Upsilon$ masses in the confined phase is based on calculations on several large quenched isotropic lattices using clover-improved Wilson valence fermions carrying different quark masses. The extrapolated lattice correlators are confronted with perturbation theory results incorporating resummed thermal effects around the threshold from potential nonrelativistic QCD (pNRQCD) and vacuum asymptotics above the threshold.
An additional transport peak is modeled below the threshold allowing for an estimate of the diffusion coefficients for charm and bottom quarks. 
We find that charmonium correlators in the vector channel can be well reproduced by perturbative spectral functions above $T_c$ where no resonance peaks for $\Jpsi$ are needed at and above 1.1 $T_c$, while for bottomonium correlators a resonance peak for $\Upsilon$ is still needed up to 1.5 $T_c$. By analyzing the transport contribution to the correlators we find that the drag coefficient of a charm quark is larger than that of a bottom quark.

\end{abstract}

\maketitle

\section{Introduction}

Heavy quark-antiquark bound states, quarkonia, have been proposed as a thermometer of quark gluon plasma in heavy ion collisions since they are formed at a very early stage of the collisions and may survive in the deep deconfined phase due to their hierarchically small sizes and large binding energies~\cite{Matsui:1986dk,Karsch:2005nk}. The suppression of quarkonia yields in AA collisions compared to those in the pp collisions have been observed in RHIC and LHC energies~\cite{Strickland:2011aa,STAR:2013eve,STAR:2013kwk,Adare:2014hje,ALICE:2014wnc,Krouppa:2015yoa,ALICE:2016flj,Sirunyan:2017lzi,CMS:2018zza,ALICE:2018wzm}, however, its interpretation is still not very clear due to the interplay between the cold and hot nuclear effects~\cite{Brambilla:2010cs}. Due to their nonperturbative features it thus is important to understand the fate of quarkonia in the hot medium from lattice QCD computations.

In addition, it was observed that open heavy mesons show an unexpectedly substantial elliptic flow that is comparable to that of light-quark mesons at RHIC~\cite{Adare:2006nq,Abelev:2006db} and LHC~\cite{ALICE:2012ab}. Moreover, heavy quarks are found to lose a significant amount of energy similar to light flavors at sufficiently high transverse momentum, while with decreasing transverse momentum an energy loss hierarchy is expected due to the dead cone effect, in short, heavier quarks suffer less energy loss. Experimentally the nuclear modification factor of open flavor mesons seems to support such a picture~\cite{Adare:2015hla,Oh:2017usm,Sirunyan:2018ktu}. Phenomenological explanations of these phenomena require a modeling of the heavy quark diffusion in a hot and dense medium. This requires knowledge about the heavy quark diffusion coefficients $D$~\cite{He:2014epa,Cao:2014fna,Cao:2018ews,Li:2021xbd} which can be determined in lattice QCD calculations as they are encoded in the correlation and spectral functions of quarkonia in the vector channel. 

In the heavy quark mass limit recent progress has been made to estimate the heavy quark momentum diffusion coefficient based on continuum extrapolated color-electric field correlation functions \cite{Francis:2015daa,Brambilla:2020siz,Altenkort:2020fgs}.
The subleading quark mass corrections to this transport coefficient are proportional to a color-magnetic field correlator \cite{Bouttefeux:2020ycy}. In the current study we will utilize full relativistic vector meson correlation functions to estimate the charm and bottom diffusion coefficients. 

The spectral functions of quarkonia in the vector channel contain all information about the in-medium hadron properties like the dissociation temperatures of the corresponding bound states and heavy quark diffusion coefficients.  However, the spectral function cannot be obtained directly from lattice QCD and is only related to lattice QCD computable Euclidean correlation functions. Investigations on quarkonium spectral functions extracted from two point correlation functions were started about two decades ago~\cite{Ding:2015ona}. Since the lattice spacing has to be smaller than the inverse of the heavy quark mass to control lattice cutoff effects and the extraction of spectral function requires a large number of data points in the temporal direction of lattices, most studies in lattice QCD focus on charmonium spectral functions and correlation functions, where continuum extrapolated results only exist for those in the pseudoscalar channel~\cite{Burnier:2017bod}. Due to the much larger mass, bottomonium spectral functions in the relativistic formalism have only been studied on highly anisotropic lattices~\cite{Liao:2001yh,Jakovac:2006sf}. Recently, studies have been carried out using nonrelativistic heavy quark formulations on anisotropic~\cite{Aarts:2014cda} and isotropic lattices~\cite{Kim:2018yhk}. A review of current status of lattice studies on heavy quarkonium in extreme conditions can be found in~\cite{Rothkopf:2019ipj, Ding:2020rtq}.

The main goal of this work is to compare lattice correlators of both charmonium and bottomonium in the vector channel with those integrated from the perturbative spectral functions. The results will be used to investigate the thermal modifications of $\Jpsi$ and $\Upsilon$ and the diffusion coefficients of charm and bottom quarks. For this we will start with the construction of perturbative spectral functions, given in Sec. \ref{sec:spf}. In Sec. \ref{sec:setup} we present the lattice setup and describe how we perform the mass interpolation and continuum extrapolation. Section \ref{sec:fit} is devoted to comparing the lattice and perturbative results in the bound state region of the vector spectral function. In Sec. \ref{sec:trans} we analyze the transport peak mainly based on a Lorentzian ansatz. In the last section we draw the conclusion. Parts of the study have been presented in various conferences and workshops~\cite{Ohno:2013rka,Ohno:2014uga,Ohno:2016ggs,Ding:2017rty, Ding:2019kva,Ding:2018uhl,Lorenz:2020uik} and in the Ph.D. thesis of Anna-Lena Lorenz \cite{Lorenz:2021}.

\section{Spectral Functions in the vector channel}
\label{sec:spf}

The quarkonium spectral function cannot be obtained directly on the lattice, and it is related to the Euclidean mesonic two point correlation function via the integral equation,
\begin{align}
\label{eq:corr_latt}
\begin{split}
G_H(\tau)&\equiv \sum_{\vec{x}}\big{\langle}\bar{\psi}\Gamma_H\psi(\tau,\vec{x})(\bar{\psi}\Gamma_H\psi(0,\vec{0}))^\dagger\big{\rangle}\\
&=\int \limits_{0}^{\infty} \frac{\dd \omega}{\pi} \rho_H(\omega)K(\omega,\tau),
\end{split}
\end{align}
where $K(\omega,\tau)=\frac{\cosh(\omega(\tau-\frac{1}{2T}))}{\sinh(\frac{\omega}{2T})}$ is a temperature ($T$) and frequency ($\omega$) dependent integration kernel. Specific to the vector channel that we are considering in this work $\Gamma_H=\gamma_\mu$ and thus $G_H(\tau)=G_{00}+G_{ii}(\tau)$. $G_{00}$ is the zeroth component independent of the distance $\tau$~\cite{Burnier:2012ts} and $G_{ii}(\tau)$ the sum of spatial components (we use the Einstein summation convention throughout this paper),
\begin{align}
\label{eq:corr_spa_comp}
G_{ii}(\tau)=\int \limits_{0}^{\infty} \frac{\dd \omega}{\pi} \rho_{ii}(\omega)K(\omega,\tau),\ \ i=1,2,3.
\end{align}

The spectral function well above the threshold and around the threshold, on the other hand, can be obtained from the perturbation theory. For frequencies well above the threshold, the spectral function can be described by ultraviolet asymptotics~\cite{Burnier:2012ts}:

\begin{align}
\rho^{vac}_{V}=\frac{3\omega^2}{4\pi}R_{c}(\omega^2)
\end{align}
with $R_c$, a polynomial in $\alpha_s$ up to five-loop order
\begin{align}
R_c(\omega^2)=&\ r_{0,0}+r_{1,0}\alpha_s + \left(r_{2,0}+r_{2,1}l\right)\alpha_{s}^2\nonumber\\
&+\left(r_{3,0}+r_{3,1}l+r_{3,2}l^2\right)\alpha_{s}^3\nonumber\\
&+\left(r_{4,0}+r_{4,1}l+r_{4,2}l^2+r_{4,3}l^3\right)\alpha_{s}^4+\mathcal{O}\left(\alpha_{s}^5\right),
\end{align}
where $l=\ln\left(\frac{\bar{\mu}^2}{\omega^2}\right)$ and $\bar{\mu}$ is the renormalization scale, whose range can be found in \cite{Burnier:2012ts}. The coefficients $r_{ij}$ for the vector channel can also be found in \cite{Burnier:2012ts}. The thermal contributions arising around the threshold can be obtained by applying pNRQCD calculations~\cite{Laine:2007gj} as
\begin{align}
\rho^{\text{pNRQCD}}_V(\omega)=\frac{1}{2}\left( 1-e^{-\frac{\omega}{T}} \right) \int\limits_{-\infty}^{\infty}\text{d}t\ e^{i\omega T}C_>(t,\vec{0},\vec{0}).
\end{align}
The threshold mentioned above is a frequency at which the free quark spectral function switches from vanishing to nonvanishing value \cite{Aarts:2005hg}. At zero momentum it locates at $2M$ with $M$ the quark mass.  $C_>$ is a Wightman function, which is solvable for a real-time static potential from hard thermal loop resummation~\cite{Laine:2006ns}. 

The two energy regimes are matched by modifying the pNRQCD result with a factor $A^{match}$, so that it smoothly connects to the vacuum asymptotics at a certain point $\omega^{match}$. This matching procedure was successfully developed in the pseudoscalar channel in \cite{Burnier:2017bod}. The resulting spectral function is valid down to frequencies around and above the threshold, and overestimates the regime $2M-\omega \ll \alpha_{s}^2 M$. An exponential cutoff $\Phi(\omega)=\theta(2M-\omega)e^{-\frac{|\omega-2M|}{T}}$ is introduced to model the spectral function for the low frequencies. The whole spectral function then reads
\begin{align}
\label{eq-pertspf}
\begin{split}
\rho_V^{pert}(\omega)=&\ A^{match}\Phi(\omega)\rho^{\text{pNRQCD}}_V(\omega)\theta(\omega^{match}-\omega)\\
&+\rho^{vac}_V(\omega)\theta(\omega-\omega^{match}).
\end{split}
\end{align}
Note that the above perturbative calculations were carried out in the Minkowski space with metric ($+$$-$$-$ $-$), where $\rho^{}_V(\omega)=\rho^{}_{ii}(\omega)-\rho^{}_{00}(\omega)$. Considering that $\rho^{}_{00}(\omega)\sim \omega\delta(\omega)$~\cite{Burnier:2012ts}, there would be no difference between $\rho^{}_{ii}(\omega)$ and $\rho^{}_V(\omega)$ around or above the threshold region. So in the following analysis we will use Eq.~(\ref{eq-pertspf}) to model $\rho^{}_{ii}(\omega)$ in this frequency region. 

As for the very low frequency region, $\rho^{}_{ii}(\omega)$ is supposed to have a transport peak. In the high temperature limit, the transport peak has the following form~\cite{Karsch:2003wy,Aarts:2005hg,Petreczky:2005nh}:
\begin{equation}
\rho_{ii}^{trans}(\omega)=3 \pi \chi_q \,\frac{T}{M}\,\omega\delta(\omega),
\end{equation}
where $\chi_q$ is the quark number susceptibility and $M$ is the quark mass. In the interacting case, the $\delta$-peak can be smeared into a Lorentzian peak with a finite width of $\eta$ (drag coefficient) as~\cite{Petreczky:2005nh} 
\begin{equation}
\rho_{ii}^{trans}(\omega)=3 \chi_q ~\frac{T}{M}\frac{\omega\eta}{\omega^2+\eta^2}.
\label{eq:Lorentzian}
\end{equation}
According to \cite{Burnier:2012ts} this expression overestimates the transport contribution for larger frequencies, so we multiply Eq.~(\ref{eq:Lorentzian}) with a cutoff function $1/\cosh(\frac{\omega}{2\pi T})$ which becomes unity as $\omega\rightarrow 0$. Applying the Einstein relation
\begin{equation}
\label{einstein}
\eta=\frac{T}{M D},
\end{equation}
one arrives at the Kubo formula which relates the spectral function and the heavy quark diffusion coefficient $D$ as
\begin{align}
\label{diff_coef}
D=\frac{1}{3\chi_q}\lim_{\omega \rightarrow 0} \frac{\rho_{ii}^{trans}(\omega)}{\omega}.
\end{align}
An estimation for the range of $D$ is possible via its relation to the heavy quark momentum diffusion coefficient $\tilde\kappa$~\cite{CaronHuot:2009uh}:
\begin{align}
D=\frac{2T^2}{\tilde\kappa}.
\label{eq:Dkappa}
\end{align}
$\tilde\kappa$ has been determined from lattice calculations before \cite{Francis:2015daa} and hints to a range of $2\pi TD \in [3.71, 6.91]$ at 1.5$T_c$. Recently a similar study in a much wider range of temperatures can be found in~\cite{Brambilla:2019oaa}. Figure ~12 of \cite{Scardina:2017ipo}, and \cite{Ding:2015ona} give an overview of different results for $2\pi TD$ for different temperatures.

\begin{table}
\centering
\begin{tabular}{|c|c|c|c|c|c|c|}
\hline
$\beta$ & $r_0/a$ & $a$[fm]($a^{-1}$[GeV]) & $N_\sigma$ & $N_\tau$ & $T/T_c$ & $\#$ confs\\ \hline
\multirow{6}{*}{7.192}& & & & 48 & 0.75 & 237\\
   & & & & 32 & 1.1 & 476\\
    & 26.6 & 0.018(11.19) & 96 & 28 & 1.3 & 336\\
    & & & & 24 & 1.5 & 336\\
    & & & & 16 & 2.25 & 237\\ \hline
\multirow{4}{*}{7.394} &\multirow{4}{*}{33.8} & \multirow{4}{*}{0.014(14.24)}& \multirow{4}{*}{120} & 60 & 0.75 & 171\\
    & & & & 40 & 1.1 & 141\\
    & & & & 30 & 1.5 & 247\\
    & & & & 20 & 2.25 & 226\\ \hline
    & & & & 72 & 0.75 & 221\\
    & & & & 48 & 1.1 & 462\\
    7.544 & 40.4 & 0.012(17.01) & 144 & 42 & 1.3 & 660\\
    & & & & 36 & 1.5 & 288\\
    & & & & 24 & 2.25 & 237\\ \hline
    & & & & 96 & 0.75 & 224\\
    & & & & 64 & 1.1 & 291\\
    7.793 & 54.1 &  0.009(22.78) & 192 & 56 & 1.3 & 291\\
    & & & & 48 &1.5 & 348\\
    & & & & 32 & 2.25 & 235\\ \hline      
\end{tabular}
\caption{The lattices with four different values of bare lattice gauge couplings $\beta$  used for the continuum extrapolation. The lattice spacing $a$ stems from Wilson-loop expectation values with $r_0=0.472(5)$~fm \cite{Sommer:2014mea}. With the relation $r_0 T_c=0.7457(45)$ from \cite{Francis:2015lha}, we obtain the temperature in units of $T_c$. On each lattice, the correlators for five to six different $\kappa$-values have been measured, see Table~\ref{tab-masses}.} 
\label{tab-latticesetup}
\end{table}

\begin{table}
\begin{tabular}{|c|c|c|c|c|c|}
\hline
  $\beta$ & $\kappa$ & $m^{}_V$[GeV] &  $\beta$ & $\kappa$ & $m^{}_V$[GeV]\\ \hline
   \multirow{7}{*}{7.192} & 0.13194 & 3.21(1)  &  & 0.132008 & 3.38(2) \\
   & 0.1315 & 3.59(1) &  & 0.1315 & 3.94(2) \\
   & 0.131 & 4.01(1) & 7.394 & 0.131 & 4.47(2) \\
   & 0.13 & 4.81(1) &  & 0.129 & 6.50(2) \\
   & 0.128 & 6.34(1)  & & 0.124772 & 10.04(1) \\
   & 0.12257 & 10.11(1) & & & \\\hline  
   \multirow{7}{*}{7.544} & 0.13236 & 3.06(2) & \multirow{7}{*}{7.793}  & 0.13221 & 3.37(1)\\
   & 0.1322 & 3.28(1) &   & 0.13209 & 3.59(1)\\
   & 0.1318 & 3.82(2) &   & 0.13181 & 4.11(1)\\
   & 0.131 & 4.86(2) &    & 0.13125 & 5.11(1)\\
   & 0.1295 & 6.70(2) &    & 0.13019 & 6.92(1)\\
   & 0.12641 & 10.23(2) &    & 0.12798 & 10.42(1)\\ \hline
\end{tabular}
\caption{Hopping parameter $\kappa$ and the corresponding ground state vector meson mass $m^{}_V$ for each lattice gauge coupling $\beta$. The ground state mass $m^{}_V$ is determined using two-state fits to the spatial correlators at 0.75 $T_c$. From this table (also visualized in Fig.~\ref{fig-masses}) it can already be seen that the obtained $m_V$ are distributed around the $J/\psi$ and $\Upsilon$ masses. \label{tab-masses}}
\end{table}

\section{Lattice Setup}
\label{sec:setup}
The spectral function described in the previous section will be compared to continuum extrapolated lattice correlators of both charmonium and bottomonium in the vector channel. To realize the large and fine lattices required for our analysis, we choose the quenched approximation. The configurations are generated with a separation of 500 sweeps each consisting of one heat bath and four overrelaxation updates. For thermalization, 2000 to 5000 warm-up sweeps have been carried out. The quarkonium correlators are measured with clover-improved Wilson fermions for five different temperatures from 0.75 to 2.25 $T_c$.\footnote{$T_c\approx313 ~$MeV as $r_0T_c=$0.7457(45)~\cite{Francis:2015lha}.} Correlators at each temperature have been computed using four different $\beta$-values. The temporal lattice extent $N_{\tau}$ varies from 48 to 96 at 0.75 $T_c$ and from 16 to 32 at the highest temperature, i.e. 2.25 $T_c$. The aspect ratio is fixed at a certain temperature, and changes from 2 to 6 from the lowest temperature to the highest temperature. The lattice sizes and the number of measured configurations are listed in Table~\ref{tab-latticesetup}. The lattice spacing is obtained using $r_0/{a}$ and with $r_0=0.472(5)$ from \cite{Sommer:2014mea}. As seen from Table~\ref{tab-latticesetup} the lattice spacings used in our simulation are sufficiently small such that both bottom and charm quarks can be accommodated on the lattice. Since the tuning of hopping parameters $\kappa$ to have the physical masses of $\Jpsi$ and $\Upsilon$ is nontrivial, five to six different values of $\kappa$ at each lattice spacing have been used to compute the correlation functions. Our lattice setup and computations thus make the interpolation of correlators to the continuum limit and the case with physical masses of $\Jpsi$ and $\Upsilon$ possible.

To compare the correlators from different lattices, they need to be renormalized. In the vector channel, there are different options regarding the renormalization. In addition to perturbative renormalization constants known up to two-loop order~\cite{Skouroupathis:2008mf,Gockeler:2010yr}, there are nonperturbatively determined renormalization constants given in \cite{Luscher:1996jn}. Another possibility is to take the continuum limit of renormalization independent ratios with the quark number susceptibility $\chi_q$ given by the zeroth component of the vector correlator $G_{00}$. Although the renormalization constants computed from~\cite{Skouroupathis:2008mf,Gockeler:2010yr} are comparable among each other, we decide on using the renormalization independent ratio, e.g. $G(\tau,T)$ divided by the quark number susceptibility for the continuum extrapolation in this work. Here we chose the value of quark number susceptibility at $T'=2.25T_c$ (denoted as $\chi'_q$) as a normalization, as the susceptibility is more precise at higher temperatures. Using continuum extrapolated results for $\chi_q^\prime/\chi_q$ at $T^\prime=2.25T_c$ and the respective temperature $T$ we obtain the correct normalization in the continuum. We also remark here that the extracted heavy quark diffusion coefficient [cf. Eq.~(\ref{diff_coef})] is also renormalization independent.

\begin{figure}[htb]
\includegraphics[width=0.48\textwidth]{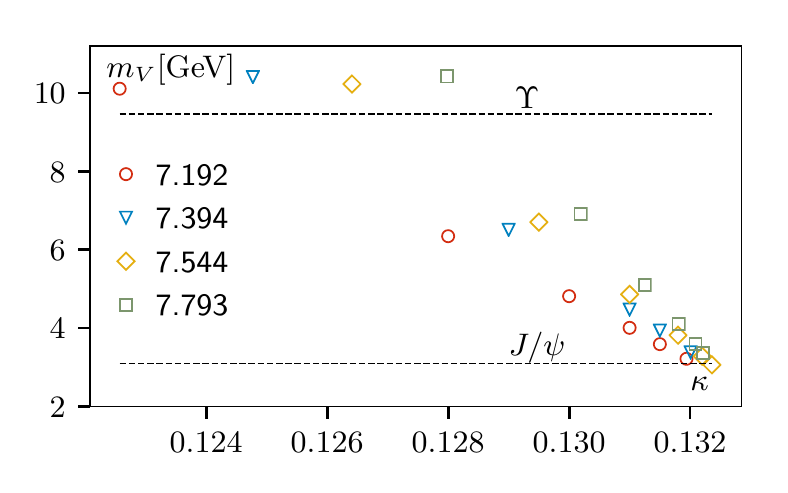}
\caption{The ground state vector meson masses $m_V$ obtained using different values of $\kappa$. The dashed lines represent the physical masses of $J/\psi$ and $\Upsilon$. It can be seen that the obtained $m_V$ from lattices are close to the physical masses, but do not exactly match. To overcome this, we interpolate the correlators between the different masses as shown in Fig.~\ref{fig-massint}. 
\label{fig-masses}}
\end{figure}

In the next step, we need to ensure that the masses and temperatures on the different lattices match. At each lattice spacing we compute the correlation functions for five to six different values of hopping parameters $\kappa$ related to the bare quark mass. The screening masses obtained at 0.75 $T_c$~\footnote{In the quenched case the screening mass obtained in the confinement phase, i.e. at 0.75 $T_c$ is supposed to be close to the pole mass of quarkonium.} shows that, due to the nontrivial quark mass tuning, the different lattices do not have the same ground state vector meson mass $m^{}_V$ (see Table~\ref{tab-masses} and Fig.~\ref{fig-masses}). To overcome this problem, an interpolation in $m^{}_V/{T}$ between the correlators computed at different values of $\kappa$ is required. We adopt the ansatz
\begin{align}\label{eq-massint}
\frac{G_{ii}\left(\tau T,\frac{m^{}_V}{T}\right)T^{\prime 2}}{T^3 \chi_q}=\exp\left(p\left(\frac{m^{}_V}{T}\right)^2 + q\frac{m^{}_V}{T} +r \right),
\end{align}
where $T'=2.25T_c$ and $(p,q,r)$ are fit parameters. 

\begin{figure}[htb]
\centering
\includegraphics[width=0.48\textwidth]{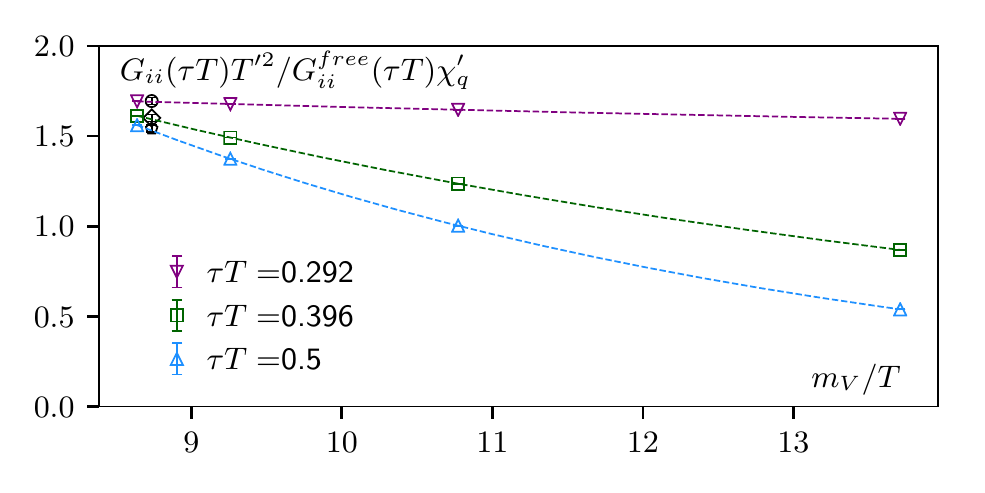}
\includegraphics[width=0.48\textwidth]{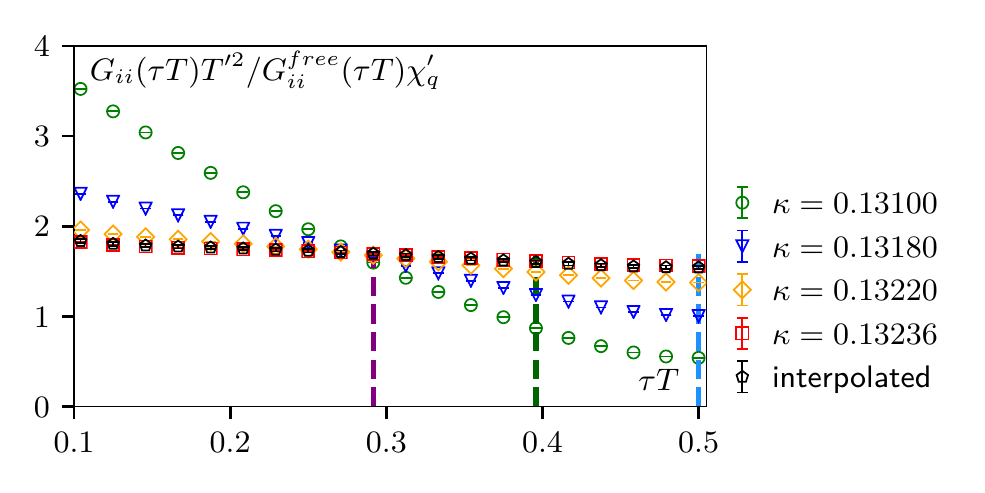}
\caption{An example of the interpolation between correlators obtained using different values of hopping parameters $\kappa$ to obtain the correlator at a physical $J/\psi$ mass on $144^3 \times 48$ lattices. From the six measured $\kappa$ values, we chose the four closest ones to the charm quark mass. For every point in $\tau T$, we interpolated with the ansatz Eq.~(\ref{eq-massint}) to the value at $m^{}_V=m_{J/\psi}$. Note that the ground state meson mass $m^{}_V$ is obtained from two-state fits to the spatial correlators at 0.75 $T_c$ and we use it for all temperatures in the mass interpolation. The top plot shows this interpolation at three example points, while the bottom plot shows the final result of the mass interpolation. \label{fig-massint}}
\end{figure}

We can see that this ansatz describes the data well as shown in the top plot of Fig.~\ref{fig-massint}. Note that for the purpose of guiding the eye, hereafter the correlators are normally shown divided by $G^{free}(\tau T)$, a correlator computed from the vector spectral function in the noninteracting case~\cite{Karsch:2003wy,Aarts:2005hg}.  They are calculated at quark mass of 1.5 GeV and 5.0 GeV for charmonium and bottomonium respectively. Then we insert physical $J/\psi$ or $\Upsilon$ mass to the fitted curve and obtain the correlators at physical mass. As an example the interpolated correlators at physical mass of $\Jpsi$ on the $144^3\times48$ lattice are shown in the bottom plot of Fig.~\ref{fig-massint}.

After the mass interpolation, we carry out a combined spline fit via which we are able to interpolate the correlators to the same points in $\tau T$ and extrapolate them to the continuum limit at the same time. We choose piecewise polynomials as an ansatz for the spline fit to our correlators,
\begin{align}
\label{combine_spline}
G_{ii}(\tau T)=\sum_{i=0}^{d} a_i\big{(}\tau T-(\tau T) _0\big{)}^i+\sum_{j=0}^{n}c_j(\tau T-t_j)_{+}^d,
\end{align} 
where
\begin{align}
(\tau T-t_j)_{+}=\left\{ \begin{array}{cl}
\ \ 0, & \tau T-t_j \leq 0,\\
\tau T-t_j, & \tau T-t_j>0.\\
\end{array} \right.
\end{align}

\begin{figure*}[thb]
\centering{
\includegraphics[width=0.48\textwidth]{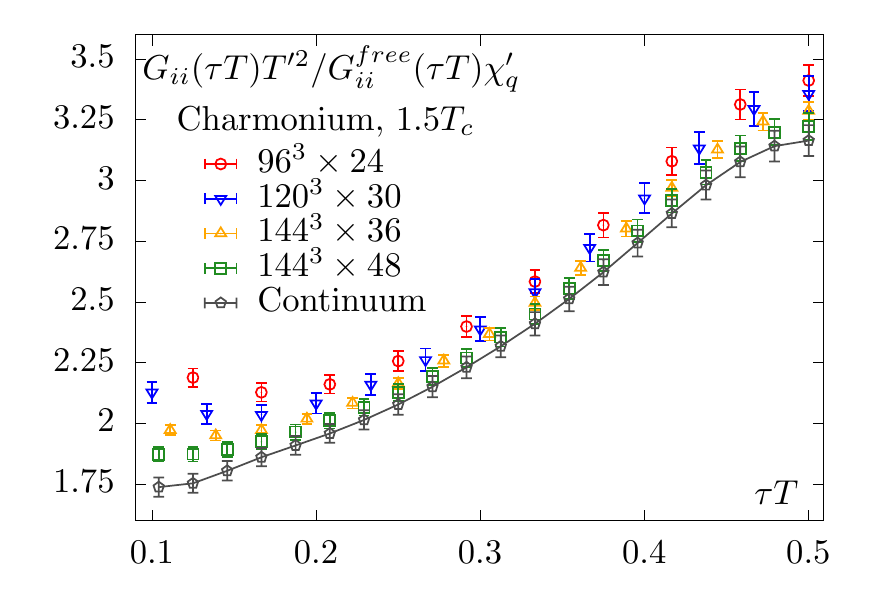}
\includegraphics[width=0.48\textwidth]{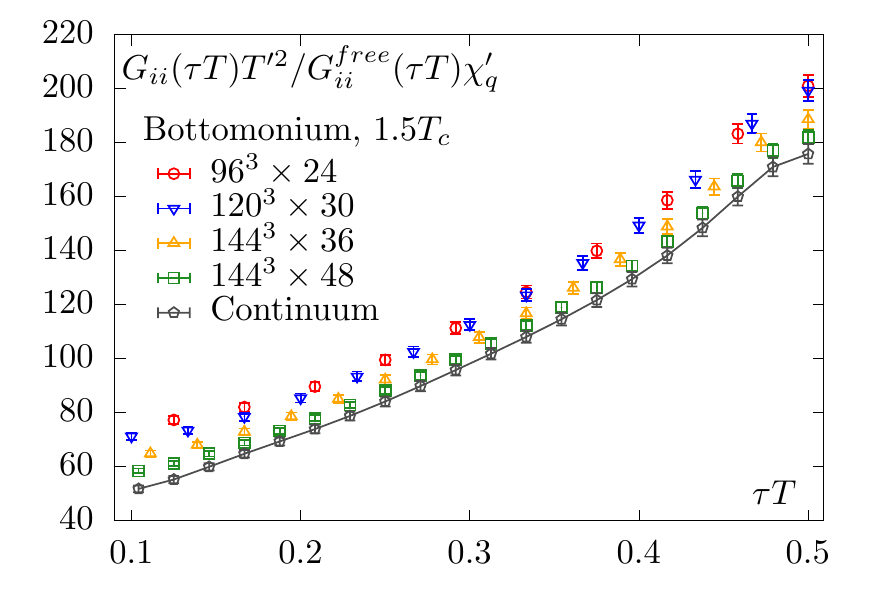}
}
\caption{The continuum extrapolation for the charmonium correlators (left) and bottomonium correlators (right) in the vector channel at 1.5 $T_c$. } 
\label{fig-contextr}
\end{figure*}
Here $d$ is the degree of the underlying polynomials, and $n$ is the number of knots which is chosen by hand for different datasets. $(\tau T)_0$ is a reference point for $\tau T$ and it cannot be any of the knots, and $t_j$ denotes the position of the knot. $a_i$ and $c_j$ are spline coefficients which can be determined when fitted to the lattice data. To incorporate lattice cutoff effects one replaces the coefficients $a_i$ with certain functions, whose form depends on how the operators concerned are constructed on the lattice. As in this work the $\mathcal{O}(a)$-improved Wilson (clover) fermions are used, one natural choice for the ansatz would be
\begin{align}
a_i=\frac{b_1}{N_\tau^2}+b_2,
\end{align}
where $b_1$ and $b_2$ are fit parameters. To obtain the continuum extrapolated values for the correlators one just needs to take $1/N^{2}_{\tau}\rightarrow 0$.

To estimate the errors, the whole procedure is conducted on bootstrap samples. We show the charmonium and bottomonium correlators on each lattice and the continuum-extrapolated correlators at 1.5 $T_c$ in Fig.~\ref{fig-contextr}. One can see that lattice cutoff effects are larger at smaller distances. Using the forementioned method we obtain a reliable continuum extrapolation down to $\tau T=0.1$. To be more certain that no cutoff effects influence our analysis, we decide to start our fits at $\tau T\approx 0.2$.
\begin{figure*}[ht]
	\centering{
		\includegraphics[width=0.48\textwidth]{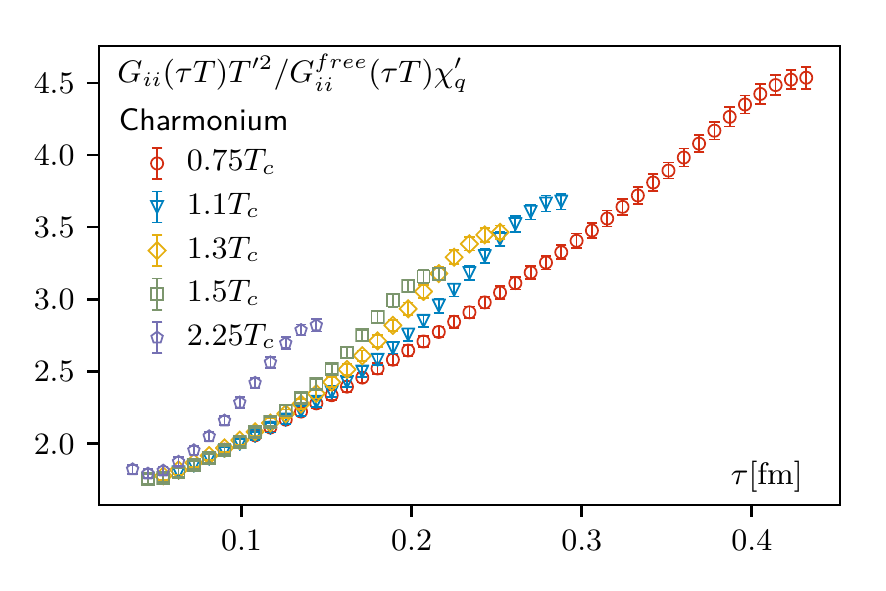}
		\includegraphics[width=0.48\textwidth]{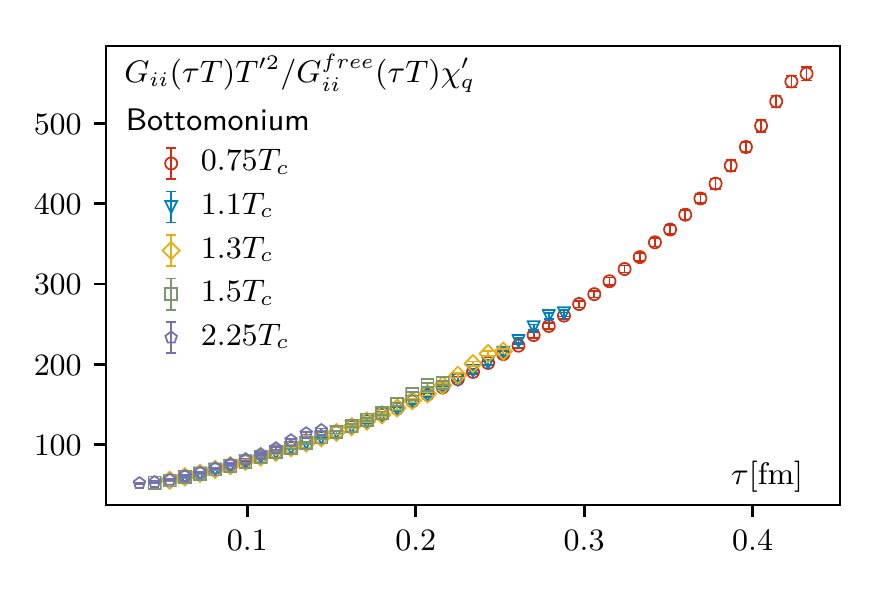}}
	\caption{The continuum extrapolated charmonium (left) and bottomonium (right) correlators divided by $G^{free}_{ii}(\tau T)\chi_q'/T'^2$ at different temperatures in the vector channel.}
	\label{fig-corrsVV}
\end{figure*}

\begin{figure*}[!htb]
\centering{
\includegraphics[width=0.45\textwidth]{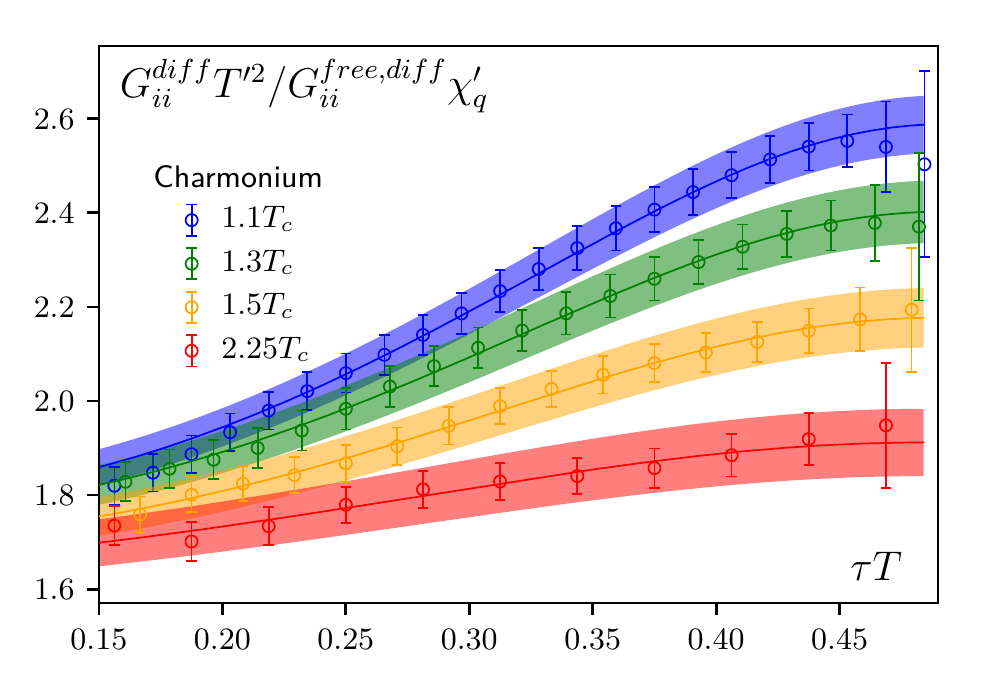}
\includegraphics[width=0.45\textwidth]{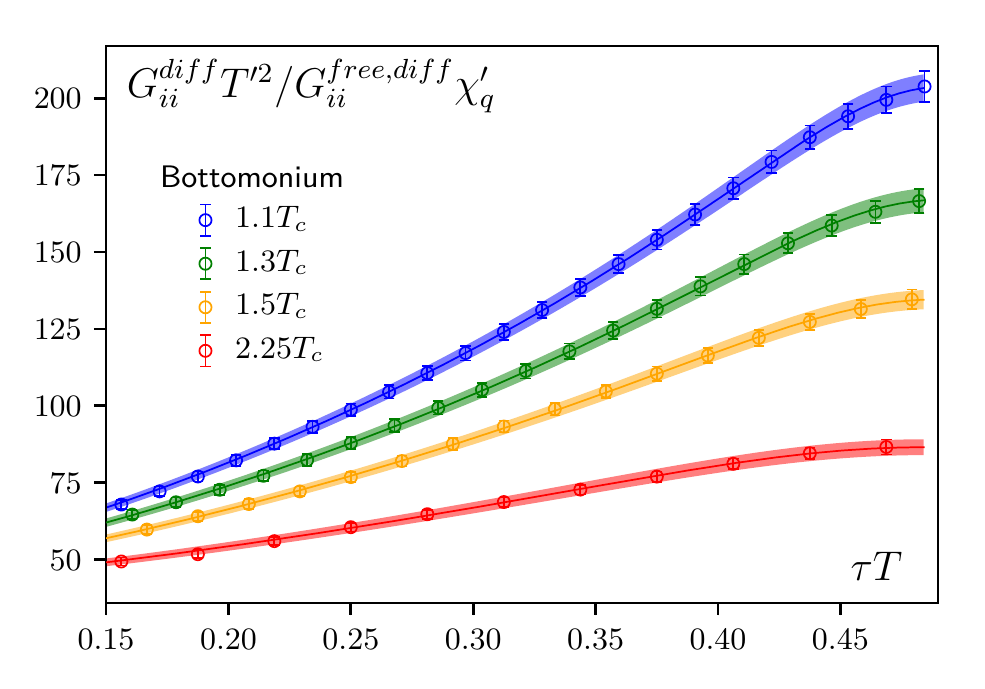}
\includegraphics[width=0.45\textwidth]{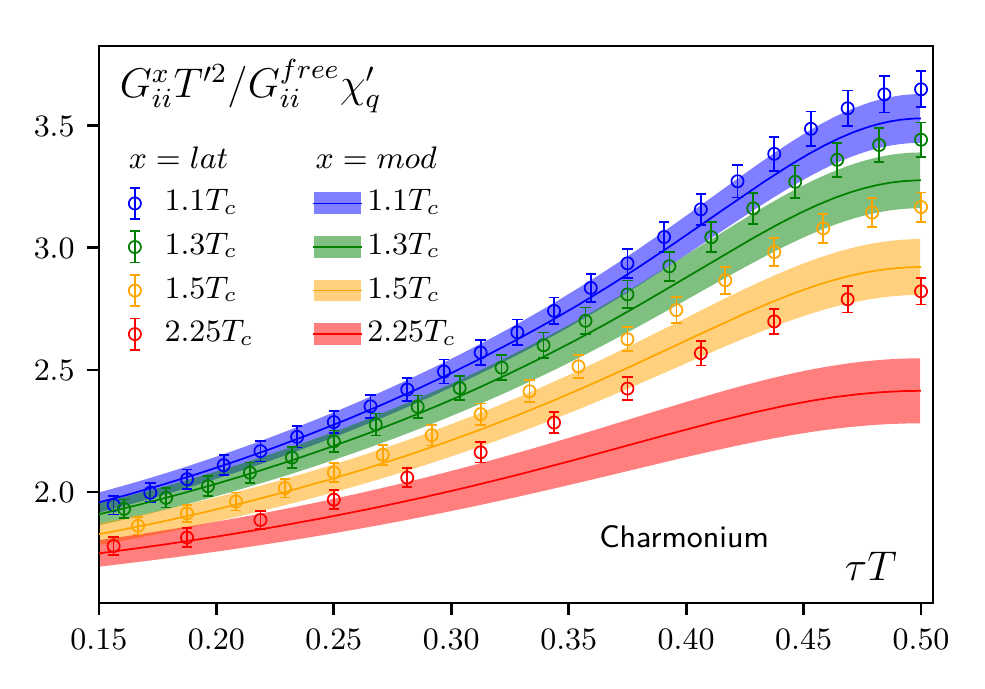}
\includegraphics[width=0.45\textwidth]{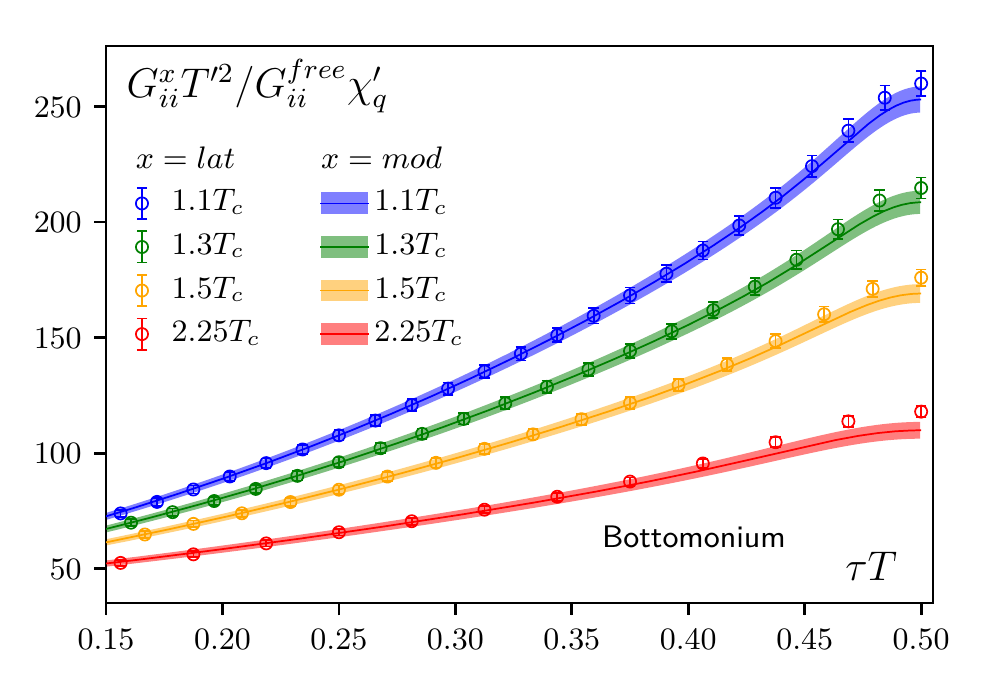}}
\caption{Fits to the difference correlator $G_{ii}^{diff}$ [cf. Eq.~(\ref{eq:Gdiff})] for charmonium (top left) and bottomonium (top right) using ansatz Eq.~(\ref{eq-rhomod}). In the bottom panel we show the original correlators ($x=lat$) and the model correlators [$x=mod$, see Eq.(\ref{full_spf_corr})] obtained from the above fits. The correlators are normalized by $G^{free,diff}_{ii}\chi_q'/T'^2$ in all cases. Here, we observe a difference between the original and model correlators that hints to a transport contribution. For bottomonium this difference is small, while it grows for charmonium
at higher temperatures.}
\label{fig:fitGdiff}
\end{figure*}

\begin{figure*}[!htb]
\centering{\includegraphics[width=0.48\textwidth]{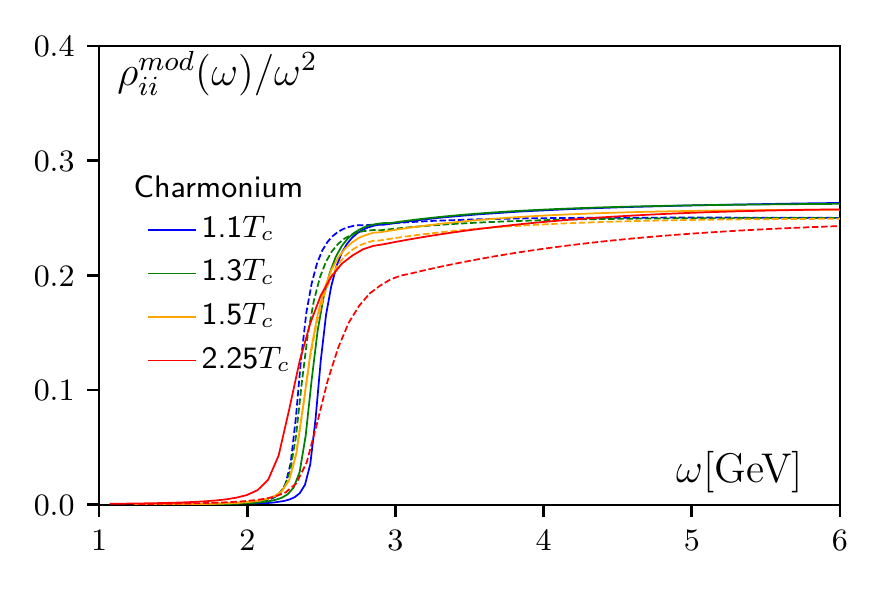}
\includegraphics[width=0.48\textwidth]{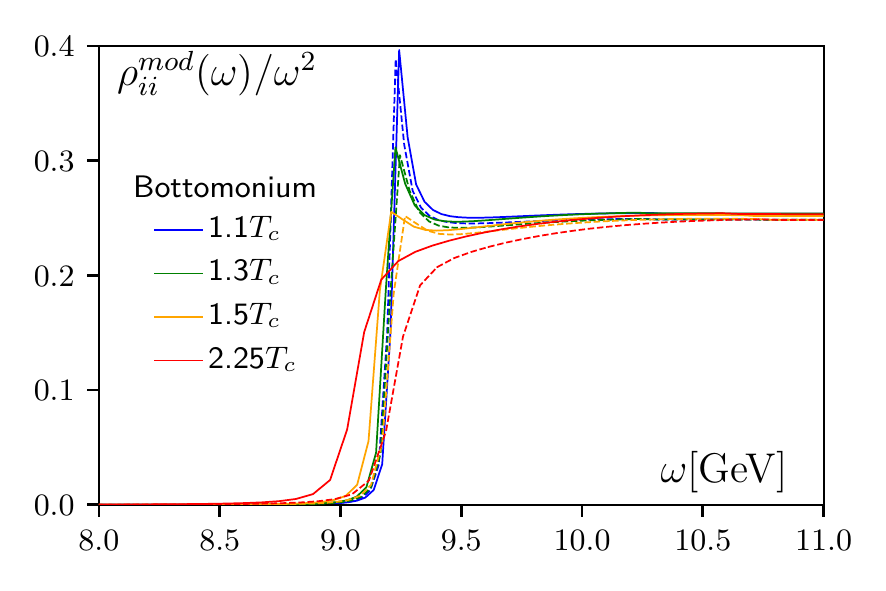}}

\centering{\includegraphics[width=0.48\textwidth]{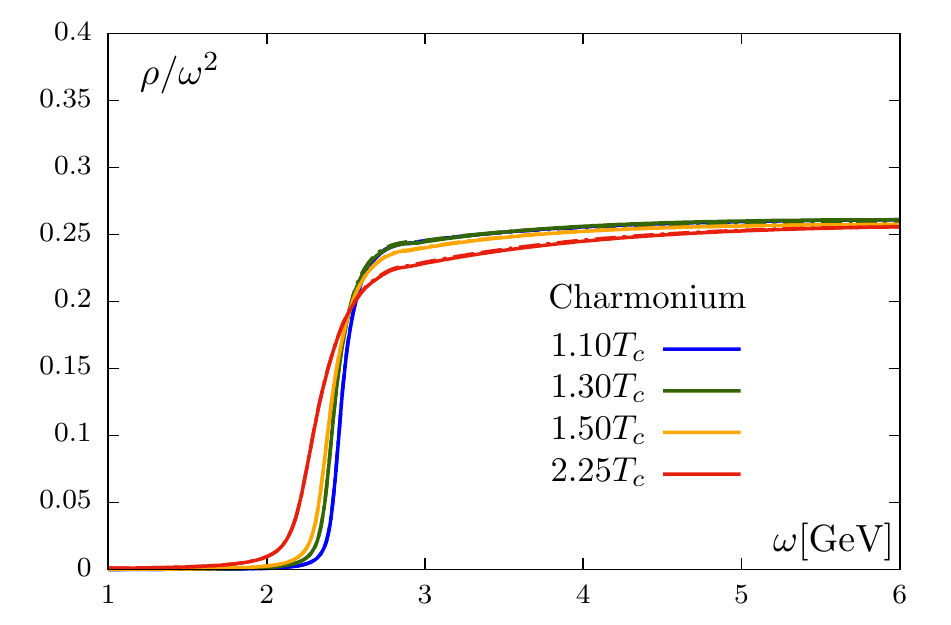}
\includegraphics[width=0.48\textwidth]{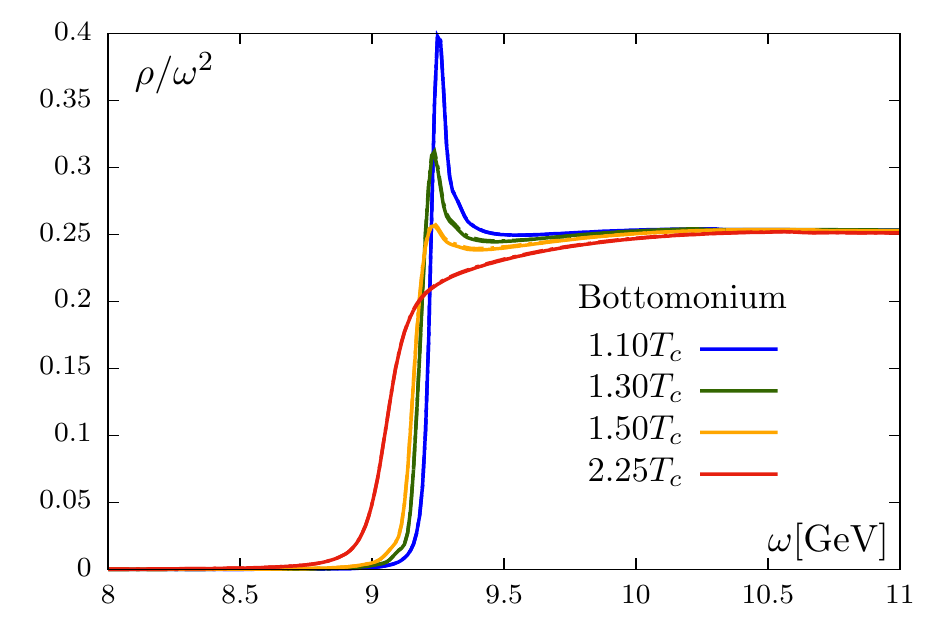}}

\caption{Top: spectral functions of charmonium (left) and bottomonium (right) in the large $\omega$ region obtained from the fits to $G_{ii}^{diff}(\tau)$. The dashed lines show the original perturbative spectral functions, while the solid lines show the modified spectral function [cf. Eq.~(\ref{eq-rhomod})]. Bottom: spectral functions for charmonium (left) and bottomonium (right) in the vector channel obtained from MEM analyses using the fit results as default models. The dashed lines are the default models and the solid lines are the outputs of MEM. The dashed lines are almost invisible as they overlap with the solid lines.}
\label{fig-spfsVVboundstate}
\end{figure*}

The final continuum extrapolated correlators at 0.75, 1.1, 1.3, 1.5 and 2.2 $T_c$ for charmonium and bottomonium in the vector channel are summarized in the left and right plots of Fig.~\ref{fig-corrsVV}, respectively. We already can draw some conclusions from the correlators without extracting the spectral functions. We see that correlators at the short distance agree for all temperatures, meaning that the region mostly influenced by the vacuum asymptotic part of the spectral function does not depend on the temperature much. At larger $\tau$, where the threshold region and the transport peak dominate the behavior, the correlators split, indicating that at least one of the two regimes is heavily temperature dependent. We can also see that comparing with bottomonium correlators, charmonium correlators have much stronger temperature dependence, especially for those at long distances which are relevant for the properties of resonance and transport peaks. This is a clear sign that charmonium suffers more thermal modifications than bottomonium.\footnote{One also sees that the ratio for bottomonium is much larger than that of charmonium. This is mostly due to the fact that the quark number susceptibility $\chi_q$ of bottom quark is much smaller than that of charm quark.}

To obtain more quantitative results, in the following section we analyze the lattice data using ansatz constructed based on the perturbative spectral function described in Sec.\ref{sec:spf}. The fits are conducted on every bootstrap sample to gain a correct error estimate. We then crosscheck the fit results by maximum entropy method (MEM) analyses. For the analyses in bootstrap a covariance matrix $C_{k,l}$ is needed. Since the bootstrap compromises the covariance matrix calculated from the continuum data, we instead use the covariance matrix of the finest lattice and rescale it to the continuum as
\begin{align}
C_{k,l}=C_{k,l}^{lat}\frac{\delta G^{cont}(\tau_k)\delta G^{cont}(\tau_l)}{\delta G^{lat}(\tau_k)\delta G^{lat}(\tau_l)},
\end{align}
where $\delta G(\tau_k)$ means the error of the correlators at distance $\tau_k$. Here the superscripts ``cont" and ``lat" stand for the continuum extrapolated and lattice results, respectively.

\section{Comparison between Lattice and Perturbative Results in the Bound State Region}
\label{sec:fit}
In \cite{Burnier:2017bod}, it was found that the perturbative spectral function was well suited for describing the lattice correlators in the pseudoscalar channel after introducing corrections for systematic errors. However, the extraction of the information on the fate of $\Jpsi$ and $\Upsilon$ from the vector correlators is more complicated since the transport peak lying in the very low frequency region is not described by the perturbative spectral function. In the following we thus divide our analyses in the two different regimes. In this section, we investigate the bound state region, i.e. the intermediate and large frequency part of the spectral function, where the perturbative spectral function is valid. The small frequency part that contains the transport peak is then evaluated in the next section. The complete spectral function and the corresponding correlator are then given by
\begin{align}
\label{full_spf_corr}
\begin{split}
&\rho_{ii}(\omega)=\rho_{ii}^{trans}(\omega)+\rho_{ii}^{mod}(\omega), \\ &G_{ii}(\tau T)=G_{ii}^{trans}(\tau T)+G_{ii}^{mod}(\tau T),
\end{split}
\end{align}
respectively. Here $\rho_{ii}^{mod}(\omega)$ is the model spectral function~\cite{Burnier:2017bod} of the form
\begin{align}\label{eq-rhomod}
\rho_{ii}^{mod}(\omega)=A\rho_V^{pert}(\omega-B).
\end{align}
The factors $A$ and $B$ correct for two sources of systematic errors that account for some of the quantitative differences in the comparison between the lattice data and perturbative results. On the lattice side, the renormalization might be off. This is taken care of by the overall normalization factor $A$. On the perturbative side, the relation between the pole mass and the $\overline{\text{MS}}$ mass is poorly determined which might lead to a slightly smaller or larger threshold location. This is taken care of by the mass shift $B$.

As the contribution from the transport peak to the correlator, i.e. $G_{ii}^{trans}(\tau)$ is nearly $\tau$ independent, one can thus look into the differences of correlators at neighboring points~\cite{Ding:2012sp}
\begin{equation}
G_{ii}^{diff}(\tau/a) = G_{ii}(\tau/a+1) -  G_{ii}(\tau/a).
\label{eq:Gdiff}
\end{equation}
In $G_{ii}^{diff}(\tau/a)$ the contribution from the transport peak, mostly influencing the correlator at $\tau T\approx 0.5$, is suppressed. In this way one can directly confront the perturbative results with the lattice data of $G_{ii}^{diff}(\tau)$ as was done in \cite{Burnier:2017bod} for correlators in the pseudoscalar channel. 

\begin{table}[!htb]
\centering
\begin{tabular}{|c||c|c||c|c|}
\hline
 & \multicolumn{2}{c||}{Charmonium} & \multicolumn{2}{c|}{Bottomonium} \\ \hline
$T/T_c$ & $A$ &  $B/T$ & $A$ &  $B/T$\\ \hline
1.1 & 1.09(2) & 0.37(4) & 1.03(2) & 0.04(2) \\
1.3 & 1.07(2) & 0.16(5) & 1.01(1) & $-0.05(2)$ \\
1.5 & 1.03(2) & 0.01(6) & 1.00(2) & $-0.12(2)$ \\
2.25& 0.99(3) & $-0.27(9)$ & 0.99(2) & $-0.23(4)$\\
\hline
\end{tabular}
\caption{Results from the fit of the model spectral function Eq.~(\ref{eq-rhomod}) to the lattice data $G_{ii}^{diff}$ [Eq.~(\ref{eq:Gdiff})] in the vector channel.}
\label{tab-AB}
\end{table}

\begin{figure*}[!thp]
	\includegraphics[width=0.47\textwidth]{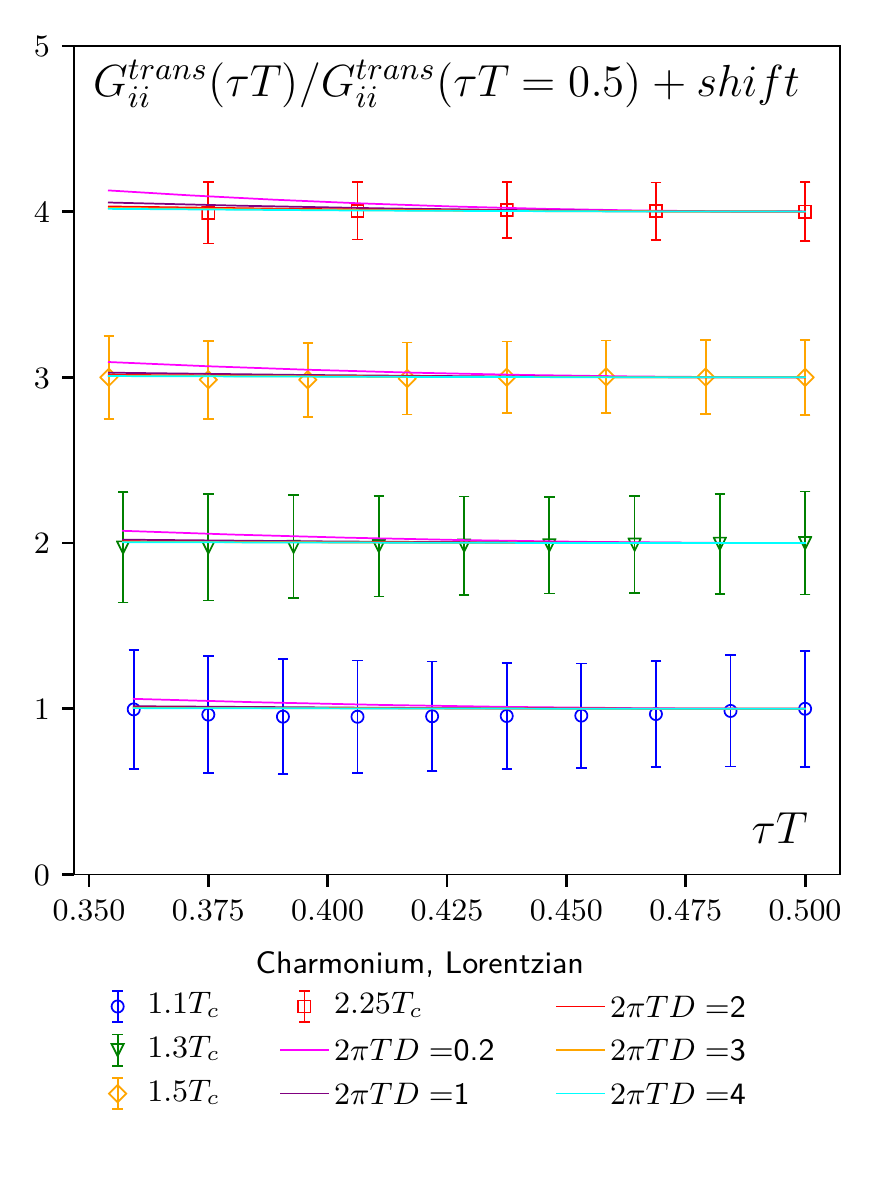}
	\includegraphics[width=0.47\textwidth]{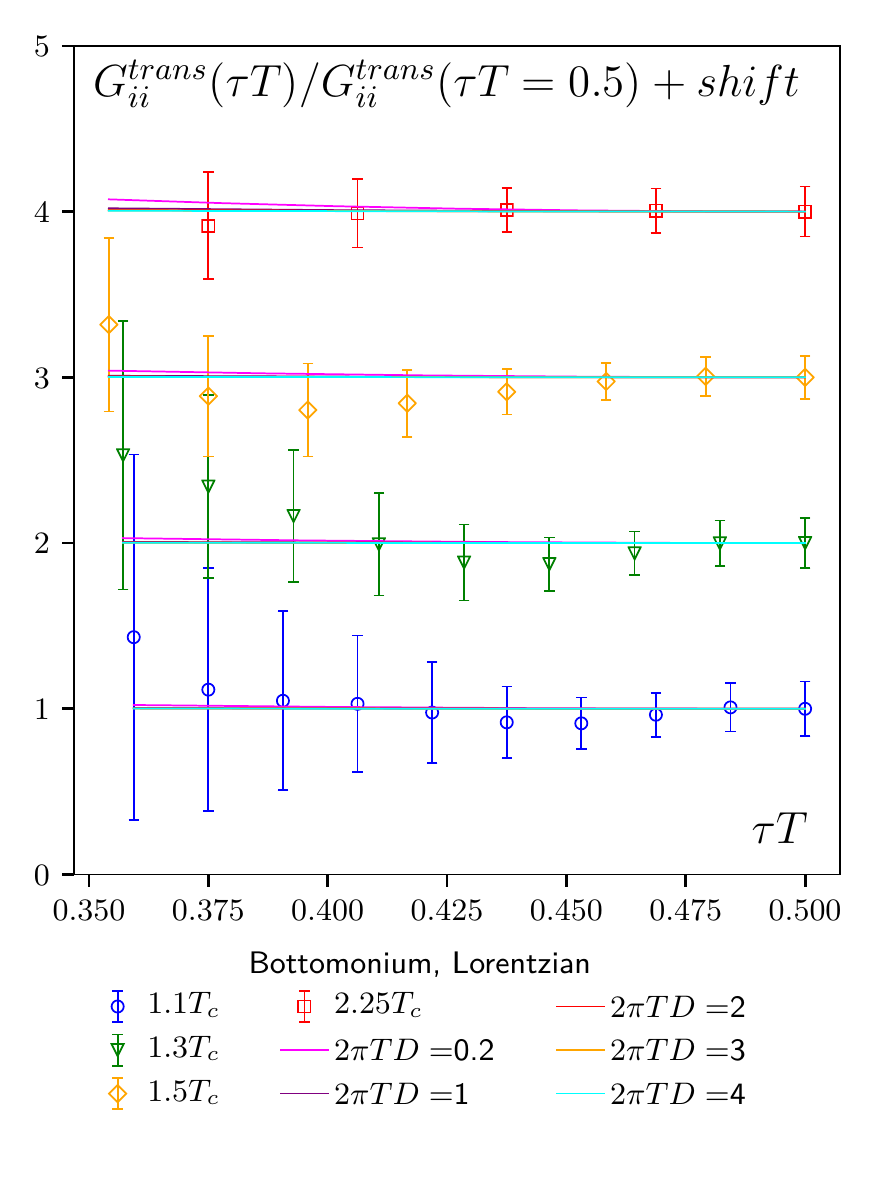}
	\caption{$G_{ii}^{trans}$ vs $\tau T$, normalized by the value at the midpoint for charmonium (left) and bottomonium (right). The points denote the lattice data at four temperatures above $T_c$, while the curves represent the corresponding results obtained using the Lorentzian ansatz [cf. Eqs.~(\ref{eq:Lorentzian} and~\ref{einstein})] for the transport peak with five different values of $2\pi TD$ at each temperature. For visibility the lattice data points and curves at $T=1.3$, 1.5 and 2.25 $T_c$ are shifted simultaneously upwards by 1, 2 and 3, respectively. It can be seen that the curvature of the correlator with the current precision does not provide any additional information to determine $2\pi TD$.}
	\label{fig:Gtrans}
\end{figure*}

Inserting the model spectral function Eq.~(\ref{eq-rhomod}) into Eq.~(\ref{eq:corr_spa_comp}), we obtain an expression for the correlator that is then fitted to the lattice data $G_{ii}^{diff}(\tau)$. The resulting parameters $A$ and $B/T$ are listed in Table~\ref{tab-AB}, and the comparison between the fits and lattice data is shown in the top panel of Fig.~\ref{fig:fitGdiff}. One can see that the lattice data is well described by the ansatz  Eq.~(\ref{eq-rhomod}). $A$ is close to one and $B$ is small, indicating that the perturbative spectral function is a suitable ansatz. The resulting spectral functions are shown in the top panel of Fig.~\ref{fig-spfsVVboundstate} for charmonium (left) and bottomonium (right). We find that there is no need for a resonance peak to describe the charmonium data in the current temperature window, while for bottomonium one thermally broadened resonance peak represents the data better at $T\leq 1.5T_c$. The position of this peak almost does not change with temperature. At 2.25 $T_c$ the peak structure is gone. As a cross-check we perform MEM analyses using the fit results as default models. The results are shown in the bottom panel of Fig.~\ref{fig-spfsVVboundstate}. We can see that for both charmonium and bottomonium at all temperatures, output spectral functions almost overlap with the inputs, which suggests the perturbative spectral function a good ansatz to describe the lattice correlators.  We remark here that the MEM analyses serve only as a consistency check, as the output spectral function from the MEM analyses is known to have large default model dependencies~\cite{Ding:2017std}.

\section{Charm and bottom quark diffusion coefficients}\label{sec:trans}

When comparing the fit results of $G_{ii}^{diff}$ to the original lattice data (see the bottom panel of Fig.~\ref{fig:fitGdiff}), we observe a difference more obviously at higher temperatures. This difference is a clear sign of a transport contribution. Qualitatively, we can already draw some conclusions, before analyzing the difference more closely in the following section. As can be seen from the bottom right plot in Fig.~\ref{fig:fitGdiff} the difference is very small for bottomonium. This indicates that the correlator is almost dominated by the bound state region. For charmonium as shown in the bottom left plot of Fig.~\ref{fig:fitGdiff}, the difference is much larger compared to the case of bottomonium and it increases with growing temperatures.

In the previous section we obtained an expression for the spectral function in the bound state region which can describe $G^{diff}_{ii}$ well. We now construct a correlator that describes the contribution from the transport peak by the subtraction
\begin{equation}\label{eq-gtrans}
G_{ii}^{trans}(\tau T) = G_{ii}(\tau T) - G_{ii}^{mod}(\tau T).
\end{equation}
The obtained $G_{ii}^{trans}(\tau T)$ for both charmonia and bottomonia at 1.1, 1.3, 1.5 and 2.25 $T_c$ are shown in the left and right plot of Fig.~\ref{fig:Gtrans}, respectively.

As expected from the findings in \cite{Petreczky:2005nh}, we observe a very weak dependence of $G^{trans}_{ii}(\tau T)$ on $\tau T$ at all temperatures, especially for the charm sector. The curvelessness of $G_{ii}^{trans}(\tau T)$ implies a slender hope to reconstruct the transport peak without further information. This is verified when we model the transport peak using a Lorentzian ansatz (see Eq.~(\ref{eq:Lorentzian})). We vary the heavy quark diffusion coefficient $D$ in the range $2\pi TD\in [0.2,4]$ and make use of the Einstein relation Eq. (\ref{einstein}) to obtain the drag coefficient $\eta$. For the values of quark masses we use $M_c=1.28$ GeV and $M_b=4.18$ GeV~\cite{Tanabashi:2018oca}. However, it is found that all the choices can describe the lattice data equally well within errors and almost no differences among $G_{ii}^{trans}(\tau T)/G_{ii}^{trans}(\tau T=0.5)$ resulting from various values of $2\pi TD$ can be seen.

\subsection{Relative magnitude of drag coefficients of charm and bottom quarks}
\label{sub:Relative}

\begin{figure}[!hbtp]
	\includegraphics[width=0.47\textwidth]{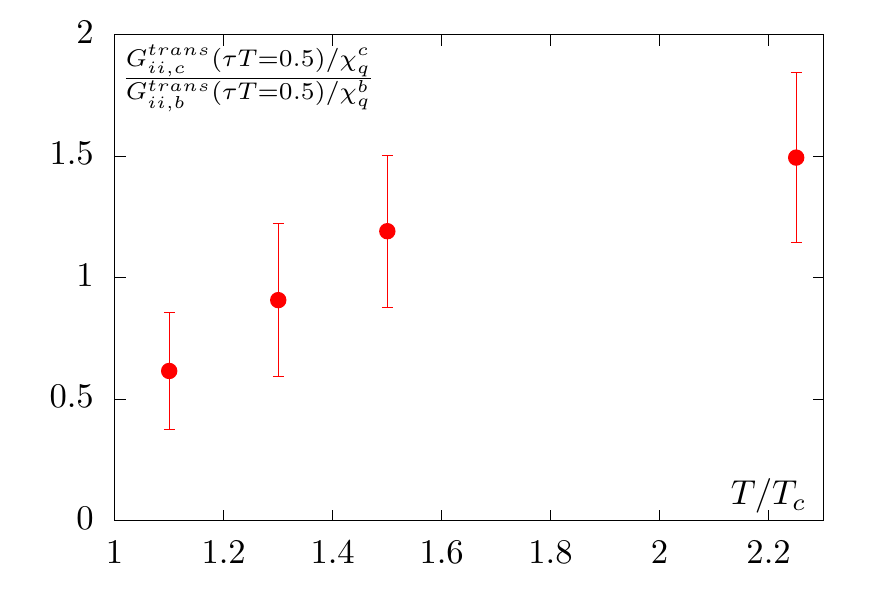}
	\includegraphics[width=0.47\textwidth]{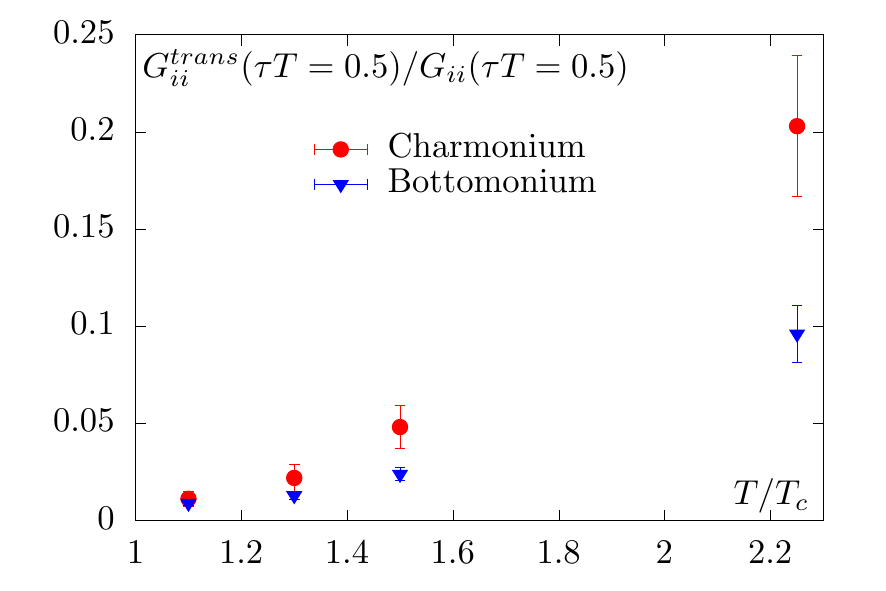}
	\caption{Top: $\frac{G_{ii,c}^{trans}/\chi_q^c}{G_{ii,b}^{trans}/\chi_q^b}$ for charmonium and bottomonium at the middle point $\tau T=0.5$ as a function of temperature. Bottom: ratio of the transport contribution to the correlator to the complete correlator. As it is seen, the transport contribution only makes up a small fraction of the correlator.}
	\label{fig:GtransMiddle}
\end{figure}

Even though $2\pi TD$ cannot be determined by analyzing the curvature of $G_{ii}^{trans}(\tau T)$ in $\tau T$, we can still draw some conclusions on the relative magnitudes of the drag coefficients of charm and bottom quark by comparing charmonium and bottomonium correlators at the midpoint ($\tau T=0.5$). The procedure to determine the relative magnitude is illustrated as follows. For small $\omega/T$, we expand the kernel and the $1/\cosh(\omega/2\pi T)$ cutoff term that is multiplied to Eq.~(\ref{eq:Lorentzian}) at the midpoint:
\begin{align}
\label{eq-expandkernel}
\frac{\cosh\left(\omega(1/2T-1/2T) \right)}{\sinh\left(\frac{\omega}{2T}\right)\cosh\left( \frac{\omega}{2\pi T} \right)}\simeq \frac{T}{\omega}\sum_{i=0}^k(-1)^{i}c_i\big{(}\frac{\omega}{T}\big{)}^{2i},
\end{align}
where $c_0=2$ and $c_1=\frac{3+\pi^2}{12\pi^2}$, for instance. With this and the Lorentzian ansatz we obtain the midpoint correlator:
\begin{align}
\begin{split}
\frac{G_{ii}^{trans}}{\chi_q T}&\simeq \frac{3T}{\pi M}\Big{[}\sum_{i=0}^k d_i\big{(}\frac{\eta}{T}\big{)}^{2i}+\sum_{i=0}^{k-1}e_i\big{(}\frac{\eta}{T}\big{)}^{2i+1}\Big{]},\\
d_i&=c_i\arctan(\frac{\omega_{cut}}{\eta}),\\ e_i&=\sum_{j=1}^{k-i}\frac{(-1)^j}{2j-1}c_{j+i}(\frac{\omega_{cut}}{T})^{2j-1}.\\
\end{split}
\end{align}
It is clear that the first term $2\arctan(\frac{\omega_{cut}}{\eta})$ is the most dominant term and higher orders are negligible for small $\eta/T$. With these simplifications, the ratio of the midpoint correlators for charmonium and bottomonium is given by
\begin{align}
\label{ratio_Gmid_chi}
\frac{G_{ii,c}^{trans}(\tau T=0.5)/\chi_{q}^c}{G_{ii,b}^{trans}(\tau T=0.5)/\chi_{q}^b}\approx \frac{M_b}{M_c}\frac{\arctan\left(\frac{\omega_{cut}}{\eta_{c}}\right)}{\arctan\left(\frac{\omega_{cut}}{\eta_{b}}\right)}.
\end{align}
As the ratio of quark masses $M_b/M_c$ is around 3~\cite{Tanabashi:2018oca}, and according to the top plot in Fig.~\ref{fig:GtransMiddle} the left-hand side of Eq.(\ref{ratio_Gmid_chi}) is even smaller than 2 at all temperatures, thus 
$\arctan\left(\frac{\omega_{cut}}{\eta_{c}}\right)/\arctan\left(\frac{\omega_{cut}}{\eta_{b}}\right)$ should be smaller than 1. Since $\arctan(1/x)$ is a monotonically decreasing function of $x$ for $x>0$, we thus have 
\begin{equation}
\eta_c > \eta_b,
\end{equation}
 i.e. the drag coefficient of a charm quark is larger than that of a bottom quark in the current temperature window. As one can also observe from the top plot of Fig.~\ref{fig:GtransMiddle} the ratio $\frac{G_{ii,c}^{trans}(\tau T=0.5)/\chi_{q}^c}{G_{ii,b}^{trans}(\tau T=0.5)/\chi_{q}^b}$ increases with increasing temperature. This could indicate that the difference between $\eta_c$ and $\eta_b$ becomes smaller at higher temperatures.

Since the curvature of $G_{ii}^{trans}(\tau T)$ can hardly provide any information on the heavy quark diffusion coefficient, in the following sections we turn to other two quantities: the midpoint correlator $G_{ii}^{trans}(\tau T=0.5)$ and the thermal moments~\cite{Ding:2012sp,Ding:2016hua,Ding:2010ga} through which there is a hope that the transport peak could be reconstructed.

\subsection{Solving transport peak using midpoint correlators}
\label{mid-trans}

\begin{figure}[!htb]
	\includegraphics[width=0.47\textwidth]{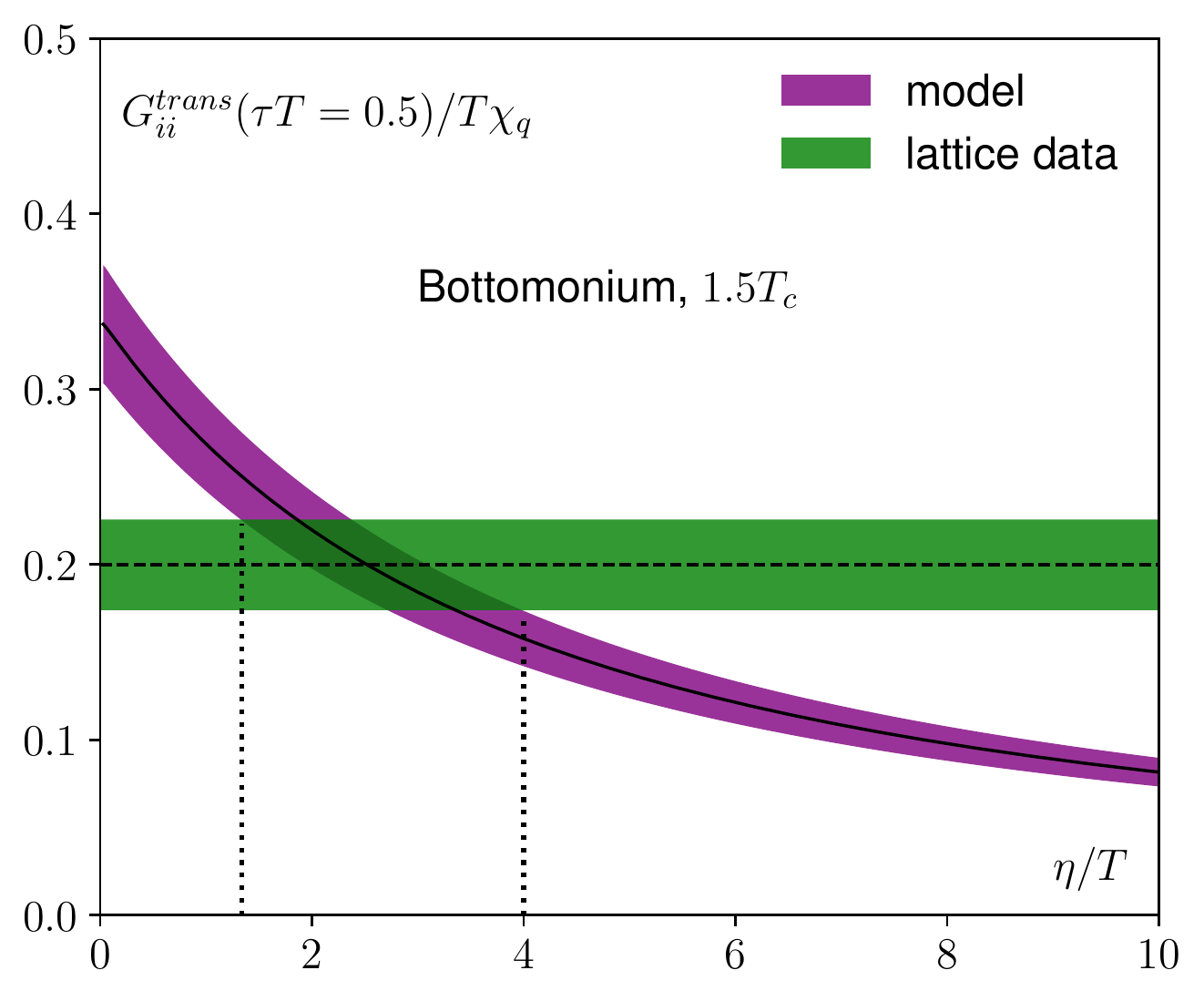}
	\caption{Estimation of $\eta/T$ for bottomonium at 1.5 $T_c$ by comparing the lattice data and the integration of ansatz (upper integration limit $\omega_{cut}/T=\infty$) at midpoint. The dashed constant line represents the mean value of the midpoint correlator $G_{ii}^{trans}$ and the band for the statistical error. The solid curve shows the integration results using a Lorentzian ansatz, and the error band is obtained by varying the quark masses $M_{c}=1.28$ GeV and $M_{b}=4.18$ GeV~\cite{Tanabashi:2018oca} by $10\%$. The intersection points (dotted) are taken as upper and lower bound of $\eta/T$.}
	\label{fig-etarangebc}
\end{figure}
In this section we consider the midpoint correlators which are shown already in Fig.~\ref{fig:GtransMiddle}. At the midpoint, the integration kernel simplifies to $1/\sinh(\frac{\omega}{2T})$. 
This allows us to compare continuum extrapolated lattice data at the midpoint to the midpoint correlator obtained in the same way as in the above section from the model spectral functions including the Lorentzian ansatz. The difference is that here we use the full kernel function and insert $M_{c}=1.28$ GeV and $M_{b}=4.18$ GeV~\cite{Tanabashi:2018oca} for the required masses and assume an error of about 10\% that accounts for the uncertainty in the definition of the mass in this approach. As an example we show the estimation of the drag coefficient of a bottom quark at 1.5 $T_c$ in Fig.~\ref{fig-etarangebc} where the middle point correlator is shown as a function of the drag coefficient. The horizontal dashed line and the surrounding green band represent the lattice data of correlator divided by $T\chi_q$ at the middle point $\tau T=0.5$, while the solid curve denotes the result obtained using the model spectral function of the Lorentzian form [cf. Eq.~(\ref{eq:Lorentzian})] and the surrounding purple band denotes the uncertainly arising from the values of quark masses. The lower and upper bound of this overlapping region between the ``lattice data" and ``model" results shown in Fig.~\ref{fig-etarangebc} can thus be regarded as the range for the estimated values of $\eta/T$. With the estimated range for $\eta/T$ we are also able to determine $D$ via the Einstein relation Eq. (\ref{einstein}). Following this procedure $\eta/T$ and $2\pi TD$ obtained for charm and bottom quarks at different temperatures are listed in Table \ref{tab-etarangebc}. 

\begin{table}[!htb]
	\centering
	\begin{tabular}{|c||c|c||c|c|}
		\hline
		& \multicolumn{2}{c||}{Charm} & \multicolumn{2}{c|}{Bottom} \\ \hline
		$T/T_c$ & $\eta/T$ & $2\pi TD$ & $\eta/T$ & $2\pi TD$\\ \hline
		1.1 & $7.37-21.38$ & $0.08-0.24$ & $<$0.81 & $>$0.66\\
		1.3 & $7.75-20.28$ & $0.10-0.26$ & $0.30-2.76$ & $0.22-2.04$\\
		1.5 & $7.93-17.08$ & $0.14-0.29$ & $1.40-4.02$ & $0.18-0.51$\\
		2.25& $4.98-10.45$ & $0.33-0.70$ & $0.62-3.20$ & $0.33-1.73$\\
		\hline
	\end{tabular}
	\caption{Estimated ranges for $\eta/T$ and the corresponding values for $2\pi TD$ according to the Einstein relation Eq. (\ref{einstein}) with a mass of $M_c=1.28\times(0.9-1.1)$ GeV and $M_{b}=4.18\times(0.9-1.1)$ GeV.}
	\label{tab-etarangebc}
\end{table}

The estimate of $\eta/T$ described above obviously depends on the upper integration limit $\omega_{cut}$. Since the transport contribution described by the Lorentzian ansatz is only valid for small $\omega$, we also investigated the effect of four different upper limits ($\omega_{cut}=\infty,M,\pi T$ and $T$) for the integration. For bottomonium, a plateau was reached, where each integration limit gave roughly the same values for $\eta/T$. Since the dependence on the integration limit was mild, we choose infinity as the upper bound. 
As for charmonium, the analysis is more complicated as the transport peak seems to be not well separated from the bound state or continuum region. Charm quark diffusion coefficient $2\pi TD$ obtained via the current approach decreases with increasing $\omega_{cut}/T$ at all the temperatures considered in the current study, and the second and third largest values of $2\pi TD$ obtained using $\omega_{cut}=M$ and $\pi T$ are almost the same, but they are at most about 1.5 times that obtained using $\omega_{cut}=\infty$ at each temperature. This might indicate that the transport peak in the charm sector has a long tail stretching to the large $\omega$ region. We thus only show the obtained values of $2\pi TD$ and $\eta/T$ for the charm quark obtained using $\omega_{cut}/T=\infty$ in Table~\ref{tab-etarangebc}. The results obtained for the charm quark thus suffer larger uncertainties than those for the bottom quark. As seen from Table~\ref{tab-etarangebc} the drag coefficient of a charm quark is larger than that of a bottom quark at each temperature (also hold for $\omega_{cut}\gtrsim \pi T$ or $M$), which is consistent with our estimate on the relative magnitude of $\eta_c$ and $\eta_b$ in the previous subsection. 

\subsection{Solving transport peak using thermal moments}

\begin{figure*}[!thb]
 \centering
    \includegraphics[width=0.47\textwidth]{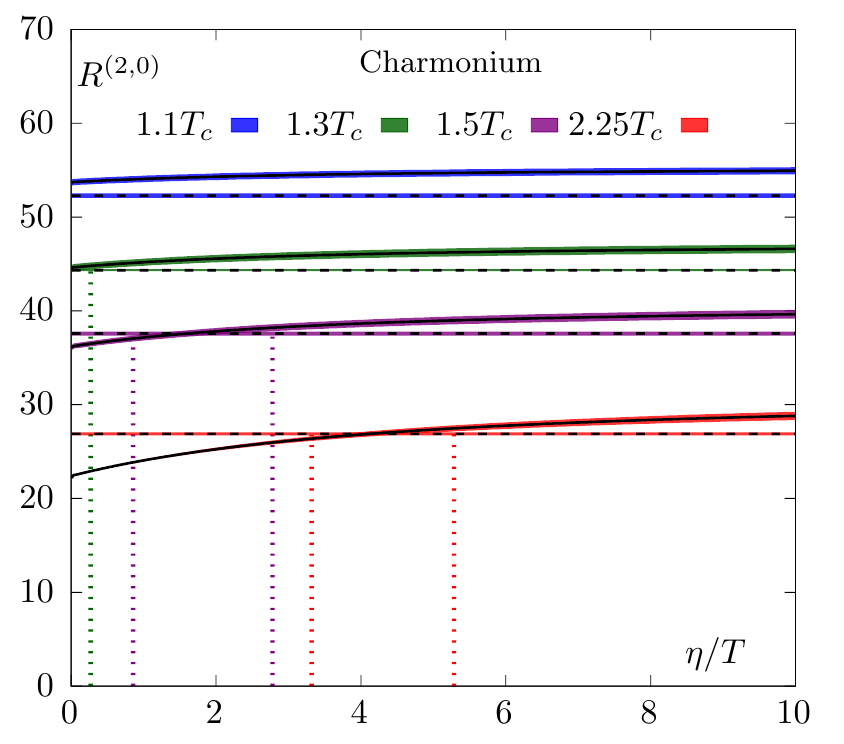}
    \includegraphics[width=0.47\textwidth]{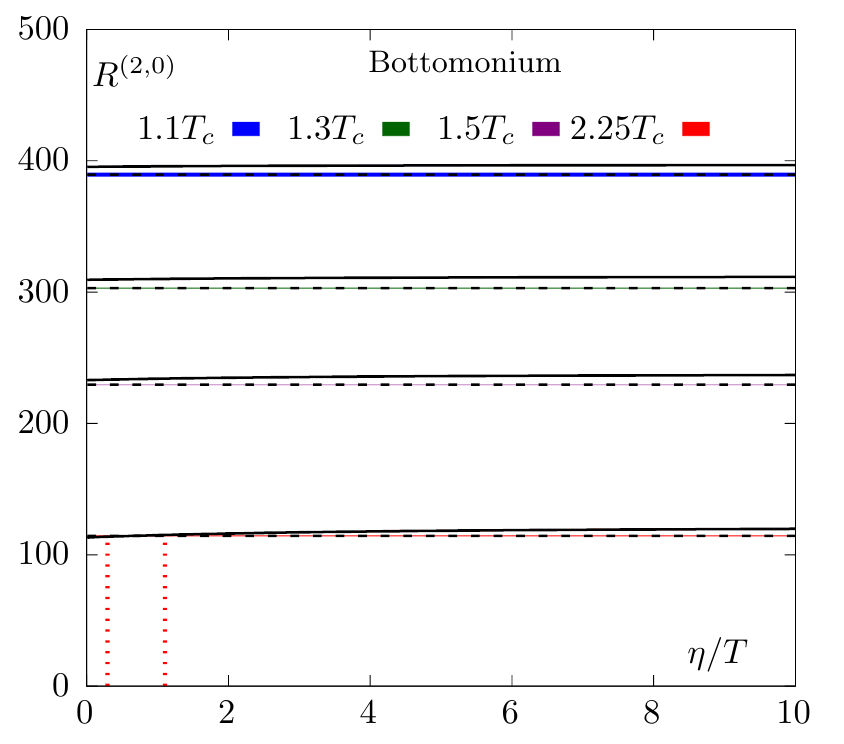}
    \caption{First thermal ratios $R^{2,0}$ for charmonium (left) and bottomonium (right) obtained from lattice computations and model spectral functions at four temperatures above $T_c$. The dashed horizontal lines are the mean values of $R^{(2,0)}$ extracted from lattice data via fits [cf. Eq.~(\ref{eq-curvfit})] and the surrounding bands denote the statistical error. The solid curves are obtained by integrating the model spectral function [cf. Eq.~(\ref{eq-r20model})] with various values of $\eta/T$ and the surrounding bands represent the uncertainty arising from the variation of quark masses by 10\%. The vertical dotted lines drawn from the intersection points between lattice and model results indicate the estimated ranges for $\eta/T$.}                    
    \label{fig-etar20}
\end{figure*}

From the above section we learn that even using the midpoint correlators it is still difficult to obtain reliable transport coefficient for charmonium. In this section we try to tackle it from the so-called thermal moments which are defined as the Taylor coefficients~\cite{Ding:2012sp,Ding:2016hua,Ding:2010ga}
\begin{align}
G_H ^{(n)}=\frac{1}{n!} \int \limits_{0}^{\infty} \frac{\dd \omega}{\pi} \left( \frac{\omega}{T} \right) ^n \frac{\rho _H (\omega)}{\sinh (\frac{\omega}{2T})}
\end{align}
when expanding the correlator around the midpoint,
\begin{align}
\begin{split}
G_H (\tau T)&=\int \limits_{0}^{\infty} \frac{\dd \omega}{\pi} \rho _H (\omega) \frac{\cosh (\omega (\tau -\frac{1}{2T}))}{\sinh (\frac{\omega}{2T})} \\
&= \int \limits_{0}^{\infty} \frac{\dd \omega}{\pi} \frac{\rho _H (\omega)}{\sinh (\frac{\omega}{2T})} \Bigg{(} 1 +\frac{1}{2!} \left( \frac{\omega}{T} \right) ^2 (\tau T -0.5)^2 \\
& \ \ \ +\frac{1}{4!} \left( \frac{\omega}{T} \right) ^4 (\tau T -0.5)^4+\cdots \Bigg{)} \\
&\approx G_{H}^{(0)} + G^{(2)}_H (\tau T-0.5)^2 + G^{(4)}_H (\tau T-0.5)^4,
\end{split}
\end{align}
where we have neglected the high order contributions in the last line. To get rid of renormalization we further build ratios
\begin{align}
R_H ^{n,m}=\frac{G_H ^{(n)}}{G_H ^{(m)}}
\end{align}
with which the expansion could be rewritten as
\begin{align}
G_H (\tau T)=G_H ^{(0)} \sum_{n=0}^{\infty} R_H ^{2n,0} (\tau T-0.5)^{2n}.
\end{align}
To obtain the moments we calculate the curvature from the data
\begin{align}
\Delta _H (\tau T)= \frac{G_H (\tau T)-G_H (\tau T=0.5)}{(\tau T-0.5)^2}.
\end{align}
The curvature could also be expressed using the ratios
\begin{align}
\label{eq-curvfit}
\frac{\Delta _H (\tau T)}{G_H (\tau T=0.5)} \approx R_H ^{2,0} \left( 1+ \sum_{n=1}^{N} R_H ^{2n+2,2n} (\tau T -0.5)^{2n} \right).
\end{align}
Now we can get the ratios $R^{2n+2,2n}_H$ by fits based on Eq.~(\ref{eq-curvfit}). Note that the approximation is valid close to the midpoint, so the fit can only be conducted on points close to $\tau T=0.5$. At the same time, to have stable fits the fit intervals cannot be too small. For this reason we vary the lower limit $\tau_{\rm{min}}T$ of the fit interval and keep the upper bound at $\tau T=0.5$. We will see that after a first few values of $\tau_{\rm{min}}T$, a plateau can be reached and we use the average over the plateau as our final estimate for $R^{2n+2,2n}$. Also for stabilities, we choose $n=1$ for charmonium and $n=2$ for bottomonium in Eq.~(\ref{eq-curvfit}).

We show the first thermal ratios obtained from fits in Fig.~\ref{fig-etar20} as horizontal constant dashed lines. To extract the information on the transport peak, we also calculate the ratios using the spectral function Eq.~(\ref{full_spf_corr}) with varying $\eta/T$. For $R^{2,0}$ we have                      
\begin{align}\label{eq-r20model}
R^{2,0}(A,B,\eta)=\frac{G^{(2)}_{mod}(A,B)+G^{(2)}_{trans}(\eta)}{G^{mod}_{ii}(\tau T=0.5)+G^{trans}_{ii}(\tau T=0.5)}
\end{align}
with
\begin{align}
\begin{split}
G^{(2)}_{mod} (A,B)=&\frac{1}{2}\int\limits_{0}^\infty \frac{\dd\omega}{\pi}\left(\frac{\omega}{T}\right)^2 A\rho_{ii}^{pert}(\omega -B)\frac{1}{\sinh\left( \frac{\omega}{2T} \right)}\,,\\
G^{(2)}_{trans}(\eta)=&\frac{1}{2}\int\limits_{0}^\infty \frac{\dd\omega}{T}\left(\frac{\omega}{T}\right)^2 3\chi_q \frac{T}{M}\frac{\omega\eta}{\omega^2+\eta^2}\\
&\times \frac{1}{\cosh\left( \frac{\omega}{2\pi T}\right)\sinh\left(\frac{\omega}{2T}\right)},
\end{split}
\end{align}
where we use $A$ and $B$ from Table~\ref{tab-AB} with statistical errors taken into account. Similarly as in the previous subsection, by searching for the intersections we manage to find a range for $\eta/T$ at some temperatures. With the Einstein relation Eq. (\ref{einstein}) a range for $2\pi TD$ could also be obtained accordingly. We list the estimates of $\eta/T$ and $2\pi TD$ for both charm and bottom quarks in Table~\ref{tab-etar20}. Our analyses using this method show that for charmonium at 1.1 $T_c$ and bottomonium at 1.1, 1.3 and 1.5 $T_c$, no intersections can be found, thus estimates for $\eta/T$ or $2\pi TD$ are not available.

\begin{table}[!thb]
\centering
\begin{tabular}{|c||c|c||c|c|}
\hline
 & \multicolumn{2}{c||}{Charm} & \multicolumn{2}{c|}{Bottom}\\ \hline
$T/T_c$ & $\eta/T$ & $2\pi TD$ & $\eta/T$ & $2\pi TD$\\ \hline
1.1 & $-$ & $-$ & $-$ & $-$\\
1.3 & $<$0.27 & $>$7.48 & $-$ & $-$\\
1.5 & $0.85-2.78$ & $0.84-2.73$ & $-$ & $-$\\
2.25& $3.32-5.28$ & $0.66-1.05$ & $0.29-1.10$ & $0.97-3.66$\\
\hline
\end{tabular}
\caption{Estimated ranges for $\eta/T$ using the thermal ratio $R^{2,0}$ and resultant $2\pi TD$ with a mass of $M_c=1.28$ GeV and $M_{b}=4.18$ GeV. For some temperatures, the method did not work out to yield a result.}
\label{tab-etar20}
\end{table}

\subsection{Combining the results on charm and bottom quark diffusion coefficients}

In previous subsections we have attempted to estimate $2\pi TD$ and $\eta$, by either analyzing the midpoint correlators $G^{trans}_{ii}$ or the thermal moments based on Lorentzian ansatz for the transport peak. It is found that when using midpoint correlators we could obtain more reliable results for the bottom quark while when using thermal moments $\eta/T$ (and also $2\pi TD$) for the charm quark is more accessible. In general all the obtained results support that  $\eta_c>\eta_b$ holds true in the current temperature window. In this section we try to combine both results by taking only the most trustworthy ones, namely 

(i) In Table~\ref{tab-etarangebc} obtained by analyzing the correlator only at the midpoint, results for bottom quark at $T\ge 1.3$ $T_c$ are chosen.

(ii) In Table~\ref{tab-etar20} obtained by analyzing the curvature of the correlator via thermal moments, results for bottom quark at 2.25 $T_c$ are chosen while those for charm quark at $T\geq1.5T_c$ are chosen.

\begin{figure}[htb]
\includegraphics[width=0.47\textwidth]{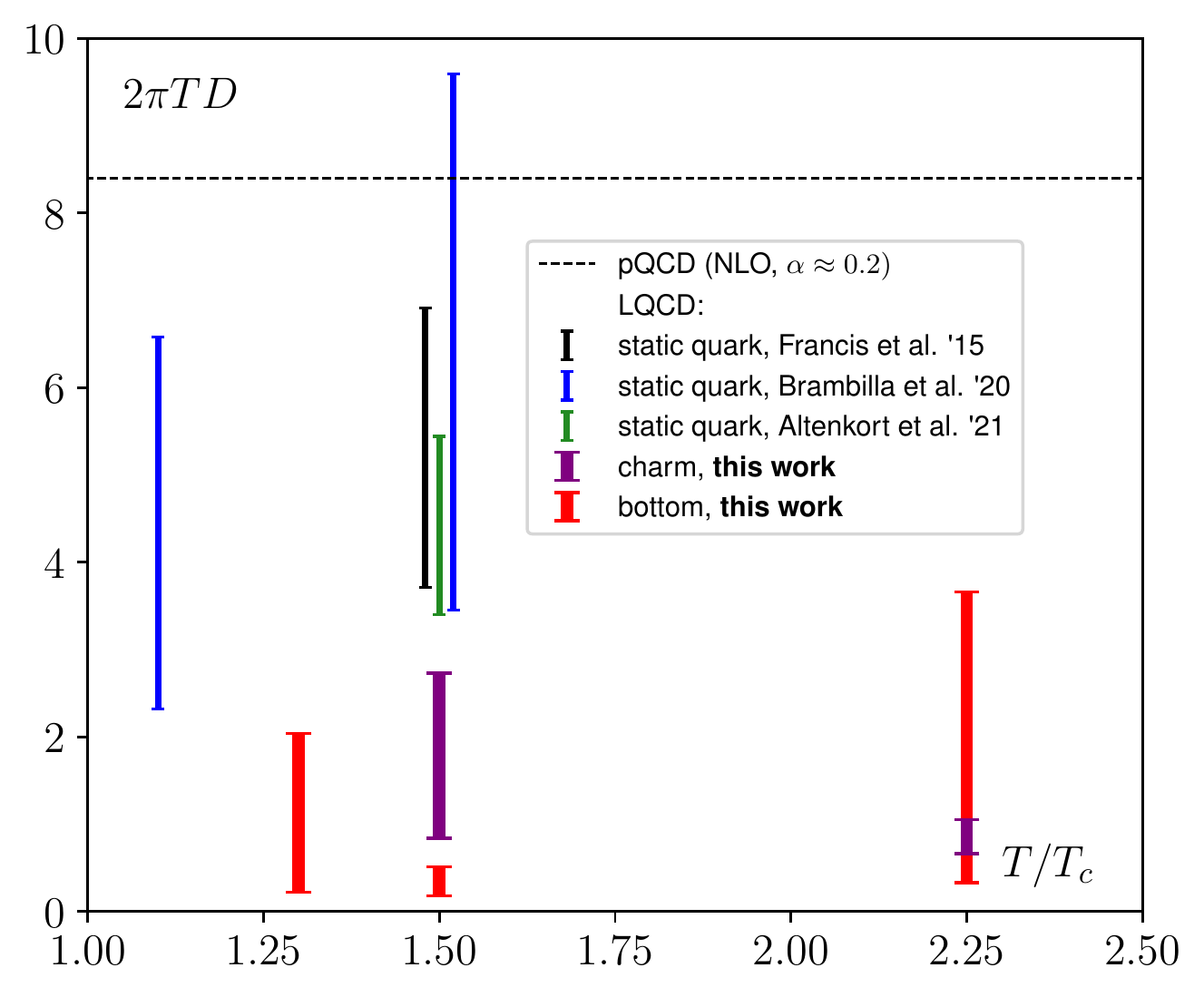}
\caption{Collection of the most reliable results for $2\pi TD$ from different analyses based on continuum extrapolated lattice results (this work). Note that the earlier results obtained by Francis $et\ al.$ 2015~\cite{Francis:2015daa}, Brambilla $et\ al.$ 2020~\cite{Brambilla:2020siz} and Altenkort $et\ al.$ 2021~\cite{Altenkort:2020fgs} are for $2\pi T D$ of a static quark rather than relativistic charm and bottom quarks considered in the current study. For comparison the results obtained from the (next-leading-order) NLO perturbative QCD with $\alpha_s\approx0.2$ (pQCD) are also shown.}
\label{fig-my2pitdrelevant}
\end{figure}

We plot the selected results in Fig.~\ref{fig-my2pitdrelevant}, as a summary of our analyses for the charm and bottom quark diffusion coefficients. In Fig.~\ref{fig-my2pitdrelevant} $2\pi TD$ for charm quark are shown at two temperatures, i.e. $T=1.5$ and 2.25 $T_c$ as red bands, while $2\pi TD$ for bottom quark are shown at three highest temperatures, i.e. $T=1.3$, 1.5 and 2.25 $T_c$ as blue bands. At 2.25 $T_c$ we have combined the estimated range for $2\pi TD$ of bottom quark obtained in Table~\ref{tab-etarangebc} and Table~\ref{tab-etar20}. We remark here that the vertical lines denote the possible ranges of the diffusion coefficients arising from the uncertainty of the heavy quark mass used in our analyses, and they do not characterize the size of the statistical error. The charm and bottom quark masses used in our analyses range from the 90\% to 110\% of their values listed by the Particle Data Group (PDG), and if we amplify the range of the heavy quark masses, the estimated range of the diffusion coefficients would become broader. For example, if we use the heavy quark masses to be 80\% to 120\% of their PDG values, the estimated range of $2\pi TD$ will become [0.14, 0.75] instead of [0.18, 0.51] in Fig.\ref{fig-etarangebc}.

As seen from Fig.~\ref{fig-my2pitdrelevant} there is no significant temperature dependence of $2\pi TD$ for both charm bottom quarks. The results of $2\pi TD$ are much smaller than those obtained in pQCD with $\alpha_s\approx0.2$. Results of $2\pi TD$ converted from the static heavy quark momentum diffusion coefficient [cf. Eq.~(\ref{eq:Dkappa})] obtained in~\cite{Francis:2015daa,Brambilla:2020siz,Altenkort:2020fgs} are also shown. These results do not show much temperature dependence as well, and are in general larger than $2\pi TD$ of both charm and bottom quarks obtained in the current study. 

We also noticed that in Refs.~\cite{Policastro:2002se, Kovtun:2003wp} a holographic estimate gives $2\pi TD=1$, but it is for R-charge diffusion. The AdS/CFT calculations for heavy quark diffusion suggest $2\pi TD=4/\sqrt{\lambda}$, where $\lambda=g^2_{YM}N_c$~\cite{Gubser:2006qh, Casalderrey-Solana:2006fio}. These estimates from AdS/CFT can be compatible with our results in this study given certain values of $g^2_{YM}N_c$.

\section{Conclusion}

In this work we have computed charmonium and bottomonium correlators in the vector channel at various quark masses on four large and fine isotropic lattices in the quenched approximation at temperatures ranging from 0.75 $T_c$ to 2.25 $T_c$. With these data we are able to interpolate the correlators to those with physical $J/\psi$ and $\Upsilon$ mass on the lattice and perform extrapolation to the continuum limit. From our analyses we see a qualitatively good agreement between our continuum extrapolated lattice data and the correlator obtained from perturbative spectral functions constructed from matching pNRQCD calculations to vacuum asymptotics. We extended the analysis in \cite{Burnier:2017bod} to the vector channel, where we divide the spectral function into the bound state region at larger $\omega$ and the transport region at small frequencies. To compare perturbation theory results and lattice data in the bound state region, we used the differences of neighboring points in the correlator and fitted a model spectral function accounting for systematic uncertainties. With this, the high frequency part is well described by the perturbative spectral function, as only mild modifications are needed. We find that for charmonium the perturbative spectral function without any resonance peak is sufficient to describe the continuum extrapolated lattice data. For bottomonium on the other hand, a thermally broadened resonance peak is needed to describe the lattice data for temperatures up to 1.5 $T_c$. Our results of spectral functions have been cross-checked using the MEM and were found to be a good description to the correlator.

For the transport contribution we analyzed the midpoint correlators and the thermal moments based on Lorentzian ansatz. We find that the drag coefficient of a charm quark is larger than that of a bottom quark. We also managed to constrain the charm and bottom quark diffusion coefficient $D$ to a possible range. Since different methods have their own (dis)advantages we combine the results by taking only the most reliable results from each method and summarize in Table~\ref{tab-combine}. We find that charm and bottom quark diffusion coefficients, obtained at physical quark mass in this study, are smaller than those converted from lattice calculations of heavy quark momentum diffusion coefficient~\cite{Francis:2015daa,Altenkort:2020fgs,Brambilla:2020siz}. The reason of such discrepancy can be that the heavy quark momentum diffusion coefficient is calculated in the heavy quark mass limit and also the studies ~\cite{Francis:2015daa,Altenkort:2020fgs,Brambilla:2020siz} only consider the leading term. Recently the subleading terms of heavy quark momentum diffusion coefficient in $T/M$ have been worked out in \cite{Bouttefeux:2020ycy} and one of them can be estimated from a color-magnetic correlator. This correlator needs to be studied in the future on the lattice and may bring the result closer to our estimates here.

\begin{table}[!thb]
\centering
\begin{tabular}{|c||c|c||c|c|}
\hline
 & \multicolumn{2}{c||}{Charm} & \multicolumn{2}{c|}{Bottom}\\ \hline
$T/T_c$ & $\eta/T$ & $2\pi TD$ & $\eta/T$ & $2\pi TD$\\ \hline
1.1 & $-$ & $-$ & $-$ & $-$\\
1.3 & $-$ & $-$ & $0.30-2.76$ & $0.22-2.04$\\
1.5 & $0.85-2.78$ & $0.84-2.73$ & $1.40-4.02$ & $0.18-0.51$\\
2.25& $3.32-5.28$ & $0.66-1.05$ & $0.29-3.20$ & $0.33-3.66$\\
\hline
\end{tabular}
\caption{Combined ranges for $\eta/T$ and $2\pi TD$ by taking the most reliable results from each method presented in Sec.\ref{sec:trans}.}
\label{tab-combine}
\end{table}

All data from our calculations, presented in the figures of this paper, can be found in~\cite{datapublication}.

\begin{acknowledgments}
We thank Rasmus Larsen and Swagato Mukherjee for the early involvement in this project and interesting discussions. This work is supported by the National Natural Science Foundation of China under Grant No. 11775096, the Guangdong Major Project of Basic and Applied Basic Research No. 2020B0301030008, the Deutsche Forschungsgemeinschaft (DFG, German Research Foundation) – Project No. 315477589 TRR 211. The computations in this work were performed on the Aachen, Bielefeld, CCNU, Juelich and Paderborn machines.
\end{acknowledgments}

\bibliography{hq}

\end{document}